\documentclass[onecolumn]{emulateapj}
\usepackage{graphicx}

\makeatletter
\newcommand*{\rom}[1]{\expandafter\@slowromancap\romannumeral #1@}

\usepackage[backref,breaklinks,colorlinks,citecolor=blue]{hyperref}
\newcommand{\OIII}{\hbox{[{\rm O}\kern 0.1em{\sc iii}]}}
\newcounter{counter}

\makeatother
\shorttitle{Cosmological constraints from Hubble parameter data}
\shortauthors{Farooq et al.}

\usepackage{verbatim}
\usepackage{graphicx}

\usepackage{float}

\usepackage{multirow}

\usepackage{threeparttable}

\usepackage{color}

\usepackage{bm}
\usepackage{tabularx}
\usepackage{cancel}
\usepackage{soul}

\begin{document}

\title{Hubble parameter measurement constraints on the redshift of the deceleration-acceleration transition, dynamical dark energy, and space curvature}

\author{Omer Farooq\altaffilmark{1}, Foram Ranjeet Madiyar\altaffilmark{1},  Sara Crandall\altaffilmark{2},  and Bharat Ratra\altaffilmark{2}}

\begin{abstract}

We compile an updated list of 38 measurements of the Hubble 
parameter $H(z)$ between redshifts $0.07 \leq z \leq 2.36$ and use them 
to place constraints on model parameters of constant and time-varying 
dark energy cosmological models, both spatially flat and curved. We use
five models to measure the redshift of the cosmological 
deceleration-acceleration transition, $z_{\rm da}$, from these $H(z)$ data.
Within the error bars, the measured $z_{\rm da}$ are insensitive to the 
model used, depending only on the value assumed for the Hubble constant 
$H_0$. The weighted mean of our measurements is $z_{\rm da} = 0.72 \pm 0.05\ (0.84 \pm 0.03)$ for $H_0 = 68 \pm 2.8\ (73.24 \pm 1.74)$ km s$^{-1}$
Mpc$^{-1}$ and should provide a reasonably model-independent estimate of 
this cosmological parameter. The $H(z)$ data are consistent with the standard 
spatially-flat $\Lambda$CDM cosmological model but do not rule out 
non-flat models or dynamical dark energy models.

\end{abstract}


\affil{$^1$Department of Physical Sciences, Embry-Riddle Aeronautical University, 600 Clyde Morris Boulevard, Daytona Beach, FL 32114, USA; farooqm@erau.edu\\
	$^2$Department of Physics, Kansas State University, 116 Cardwell Hall, Manhattan, KS 66506, USA}

\section{Introduction}

In the standard scenario the currently accelerating cosmological 
expansion is a consequence of dark energy dominating the current 
cosmological energy budget; at earlier times non-relativistic 
(cold dark and baryonic) matter dominated the energy budget and
powered the decelerating cosmological expansion.\footnote{For reviews of this picture, as well as of the alternate modified 
gravity scenario, see \cite{RatraVogeley2008}, 
\cite{Weinbergetal2013}, \cite{Martin2012}, \cite{Joyceetal2016},
and references therein.} Initial quantitative observational support for this picture came from ``lower'' redshift Type Ia supernova (SNIa) apparent magnitude 
observations and ``higher'' redshift cosmic microwave background (CMB) 
anisotropy measurements.

More recently, cosmic chronometric and baryon acoustic oscillation 
(BAO) techniques \citep[see, e.g.,][]{Simonetal2005, Morescoetal2012,
Buscaetal2013} have resulted in the measurement of the cosmological 
expansion rate or Hubble parameter, $H(z)$, from the present epoch
back to a redshift $z$ exceeding 2, higher than currently probed by SNIa
observations. This has resulted in the first mapping 
out of the cosmological deceleration-acceleration transition, the epoch
when dark energy took over from non-relativistic matter, and the first 
measurement of the redshift of this transition \citep[see, e.g.,][]{FarooqRatra2013b, Farooqetal2013b, 
Morescoetal2016}.\footnote{See \cite{SutherlandRothnie2015} and \cite{MuthukrishnaParkinson2016} for 
lower limits on this redshift derived using SNIa and other data. For upper 
limits on the transition redshift see \cite{Ranietal2015}.}

$H(z)$ measurements have also been used to constrain some more 
conventional cosmological parameters, such as the 
density of dark energy and the density of non-relativistic matter 
\citep[see, e.g.,][]{SamushiaRatra2006, ChenRatra2011b, FarooqRatra2013a, 
Akarsuetal2014, ChimentoRicharte2013, GruberLuongo2014, Bambaetal2014, Ferreiraetal2013,
Forte2014, Chenetal2015, Dankiewiczetal2014, Capozzielloetal2014, 
Mengetal2015, GuoZhang2016, MukherjeeBanerjee2016, Alametal2016}, 
typically providing constraints comparable to or better than 
those provided by SNIa data, but not as good as those from BAO or CMB 
anisotropy measurements. More recently, $H(z)$ data has been used to 
measure the Hubble constant $H_0$ \citep[][]{Verdeetal2014, Chenetal2016a},
with the resulting $H_0$ value being more consistent with recent lower values
determined from a median statistics analysis of Huchra's $H_0$ compilation
\citep[][]{ChenRatra2011a}, from CMB anisotropy data \citep[][]{Hinshawetal2013, Sieversetal2013, Adeetal2015}, from BAO 
measurements \citep[][]{Aubourgetal2015, Rossetal2015, LHuillierShafieloo2016}, and from current cosmological data and the standard model of particle 
physics with only three light neutrino species \citep[see, e.g.,][]{Calabreseetal2012}.

In this paper, we put together an updated list of $H(z)$ measurements, 
compared to that of \cite{FarooqRatra2013b}, and use this compilation 
to constrain the redshift of the cosmological deceleration-acceleration transition, 
$z_{\rm da}$, as well as other cosmological parameters. In the 
$z_{\rm da}$ analysis here we study more models than used by 
\cite{FarooqRatra2013b} and \cite{Farooqetal2013b}, now also allowing 
for non-zero spatial curvature in the XCDM parametrization of dynamical 
dark energy case and in the dynamical dark energy $\phi$CDM model 
\citep{Pavlovetal2013}. The cosmological parameter constraints derived here are based on more, as well
as more recent, $H(z)$ data than were used by \cite{Farooqetal2015} and we 
also explore a much larger range of parameter space in the non-flat 
$\phi$CDM model than they did.


\begin{table*}
\begin{center}
\begin{threeparttable}
\caption{Hubble parameter versus redshift data}
\begin{tabular}{cccc}
\hline\hline
\multirow{2}{*}{~~$z$} & ~~$H(z)$ &~~~~~~~ $\sigma_{H}$ &\multirow{2}{*}{~~ Reference\tnote{a}}\\
~~~~~    & (km s$^{-1}$ Mpc $^{-1}$) &~~~~~~~ (km s$^{-1}$ Mpc $^{-1}$)& \\
\tableline\\*[-4pt]

0.070&~~	69&~~~~~~~	19.6&~~	5\\

0.090&~~	69&~~~~~~~	12&~~	1\\

0.120&~~	68.6&~~~~~~	26.2&~~	5\\

0.170&~~	83&~~~~~~~	8&~~	1\\

0.179&~~	75&~~~~~~~	4&~~	3\\

0.199&~~	75&~~~~~~~	5&~~	3\\

0.200&~~	72.9&~~~~~~	29.6&~~	5\\

0.270&~~	77&~~~~~~~	14&~~	1\\

0.280&~~	88.8&~~~~~~	36.6&~~	5\\

0.352&~~	83&~~~~~~~	14&~~	3\\

0.380&~~	81.5&~~~~~	1.9&~~	10\\

0.3802&~~	83&~~~~~	13.5&~~	9\\

0.400&~~	95&~~~~~~~	17&~~	1\\

0.4004&~~	77&~~~~~	10.2&~~	9\\

0.4247&~~	87.1&~~~~~	11.2&~~	9\\

0.440&~~	82.6&~~~~~	7.8&~~	4\\

0.4497&~~	92.8&~~~~~	12.9&~~	9\\

0.4783&~~	80.9&~~~~~	9&~~	9\\

0.480&~~	97&~~~~~~~	62&~~	2\\

0.510&~~	90.4&~~~~~	1.9&~~	10\\

0.593&~~	104&~~~~~~	13&~~	3\\

0.600&~~	87.9&~~~~~	6.1&~~	4\\

0.610&~~	97.3&~~~~~	2.1&~~	10\\

0.680&~~	92&~~~~~~~	8&~~	3\\

0.730&~~	97.3&~~~~~	7&~~	4\\

0.781&~~	105&~~~~~~	12&~~	3\\

0.875&~~	125&~~~~~~	17&~~	3\\

0.880&~~	90&~~~~~~~	40&~~	2\\

0.900&~~	117&~~~~~~	23&~~	1\\

1.037&~~	154&~~~~~~	20&~~	3\\

1.300&~~	168&~~~~~~	17&~~	1\\

1.363&~~	160&~~~~~~	33.6&~~	8\\

1.430&~~	177&~~~~~~	18&~~	1\\

1.530&~~	140&~~~~~~~	14&~~	1\\

1.750&~~	202&~~~~~~~	40&~~	1\\

1.965&~~	186.5&~~~~~	50.4&~~	8\\

2.340&~~	222&~~~~~~~	7&~~	7\\

2.360&~~	226&~~~~~~~	8&~~	6\\ [2pt]

\hline\hline

\label{table:Hzdata}

\end{tabular}

\begin{tablenotes}

\item[a]{Reference numbers:
1.\ \cite{Simonetal2005}, 2.\ \cite{Sternetal2010}, 
3.\ \cite{Morescoetal2012}, 4.\ \cite{Blakeetal2012},
5.\ \cite{Zhang2012}
6.\ \cite{FontRiberaetal2014}, 7.\ \cite{Delubacetal2015}, 
8.\ \cite{Moresco2015}, 9.\ \cite{Morescoetal2016},
10.\ \cite{Alam2016}.

}

\end{tablenotes}
\end{threeparttable}
\end{center}
\end{table*}


We find, from the likelihood analyses, that the $z_{\rm da}$ values
measured from the $H(z)$ data agree within the error bars in all five 
models. They, however, depend more sensitively on the value of $H_0$ 
assumed in the analysis. These results are consistent with those found
in \cite{FarooqRatra2013b} and \cite{Farooqetal2013b}. In addition,
the binned $H(z)$ data in redshift space show qualitative visual 
evidence for the deceleration-acceleration transition, independent of 
how they are binned provided the bins are narrow enough, in agreement
with that originally found by \cite{Farooqetal2013b}. Given that the 
measured $z_{\rm da}$ are relatively model independent, it is not
unreasonable to average the measured values to determine a reasonable
summary estimate. We find, for a weighted mean estimate, $z_{\rm da} = 0.72 \pm 0.05\ (0.84 \pm 0.03)$ if we assume 
$H_0 = 68 \pm 2.8\ (73.24 \pm 1.74)$ km s$^{-1}$ Mpc$^{-1}$.

The constraints on the more conventional cosmological parameters,
such as the density of dark energy, derived from the likelihood
analysis of the $H(z)$ data here, indicate that these data are 
quite consistent with the spatially-flat $\Lambda$CDM model, the 
standard model of cosmology where the cosmological constant $\Lambda$ is
the dark energy. These $H(z)$ data, however, do not rule out the possibility
of dynamical dark energy or space curvature, especially when included
simultaneously, in agreement with the conclusions of \cite{Farooqetal2015}.
Currently available SNIa, BAO, growth factor, CMB anisotropy, and other 
data can tighten the constraints on these parameters, and it will be 
interesting to study these data sets in conjunction with the $H(z)$
data we have compiled here, but this is beyond the scope of our
paper. Near-future data will also result in interesting limits
\citep[see, e.g.,][]{Podariuetal2001a, Pavlovetal2012, Basseetal2014, 
Santosetal2013}.

The outline of our paper is as follows. In the next section, we discuss 
and tabulate our new $H(z)$ data compilation. In Sec.\ 3 we summarize how 
we bin the $H(z)$ data in redshift space and list binned $H(z)$ data. 
Section 4 summarizes the cosmological models we consider. In Sec.\ 5 we 
discuss how we compute and measure the deceleration-acceleration transition 
redshift and tabulate numerical values of $z_{\rm da}$ determined from the 
$H(z)$ measurements. Section 6 presents the constraints on cosmological 
parameters, and we conclude in the last section.

\section{New Hubble parameter data compilation}

\begin{figure}[H]
\centering
  \includegraphics[width=150mm]{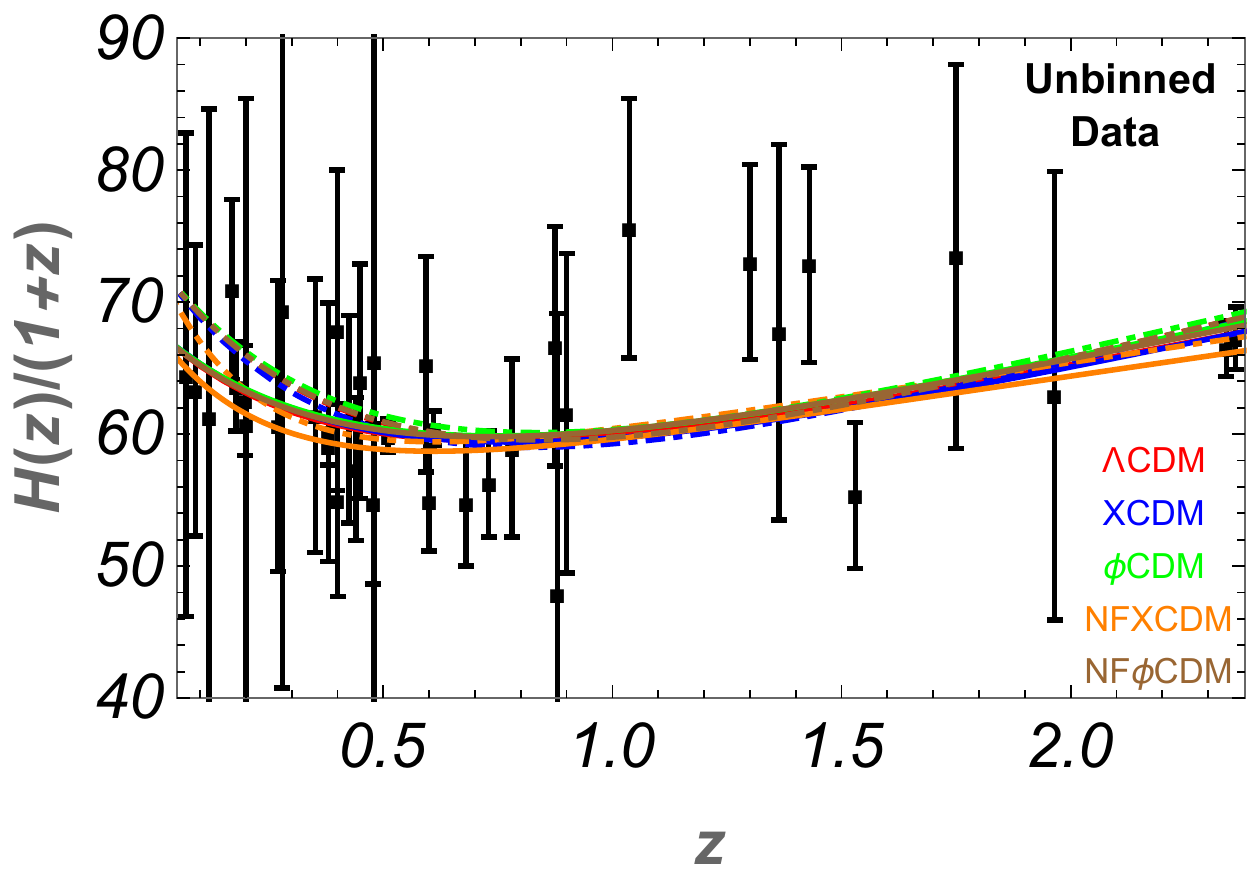}
  \includegraphics[width=88mm]{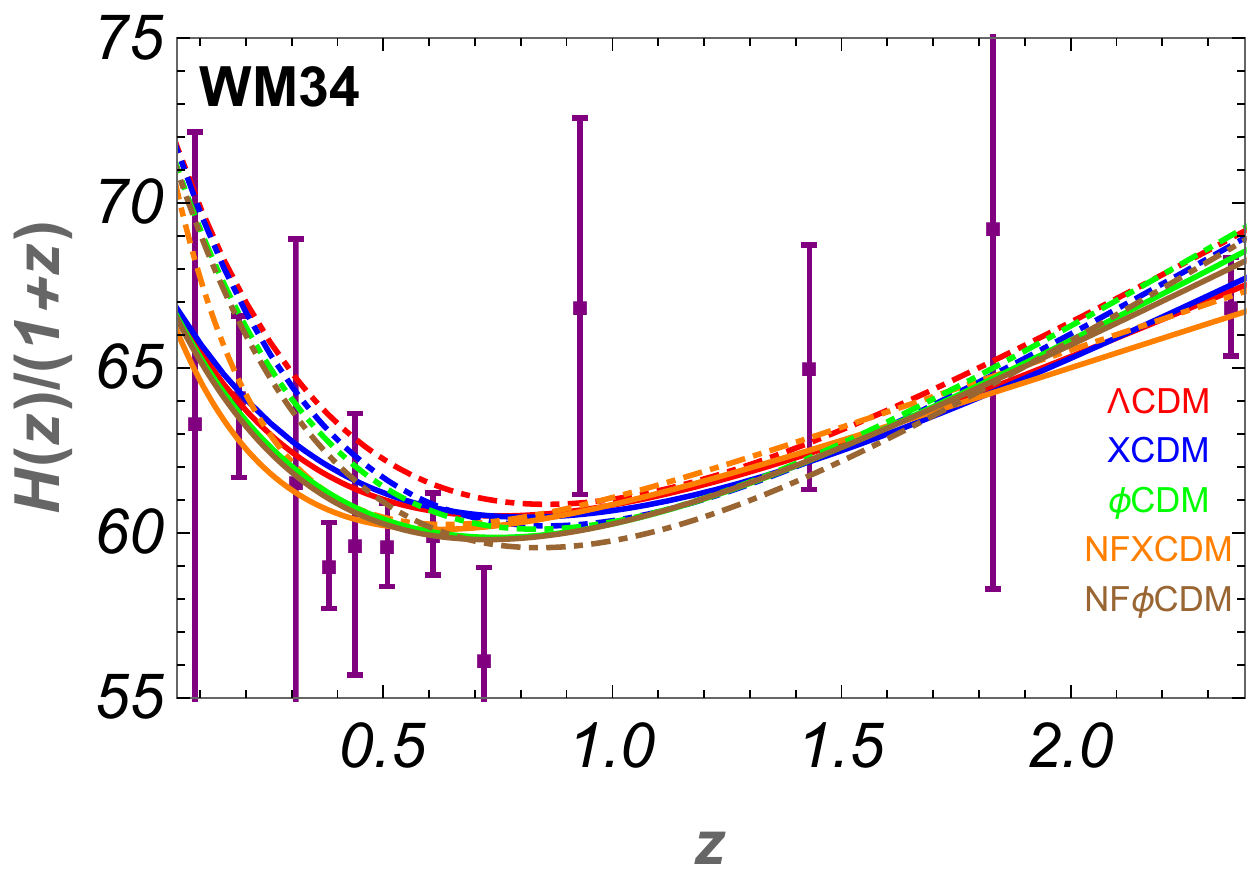}
  \includegraphics[width=88mm]{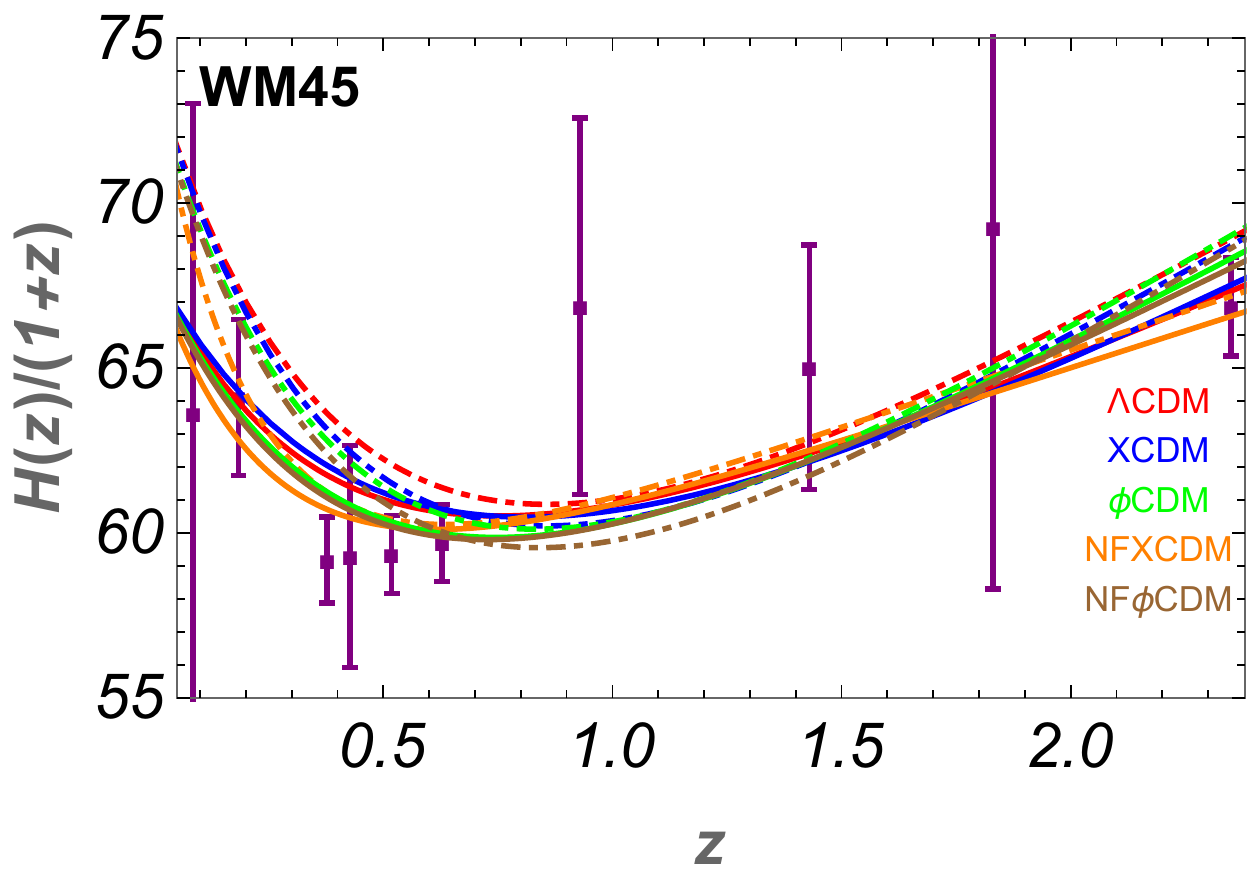}
  \includegraphics[width=88mm]{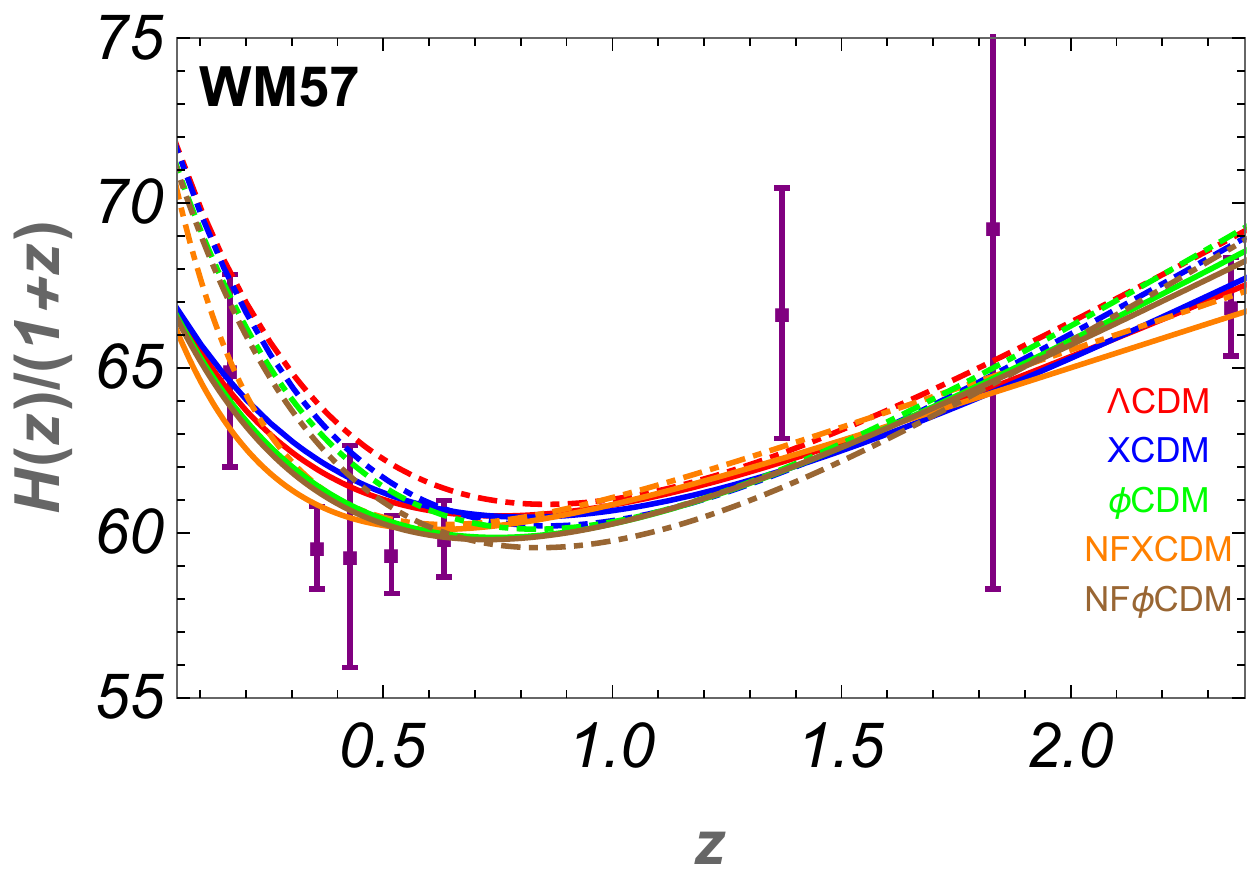}
  \includegraphics[width=88mm]{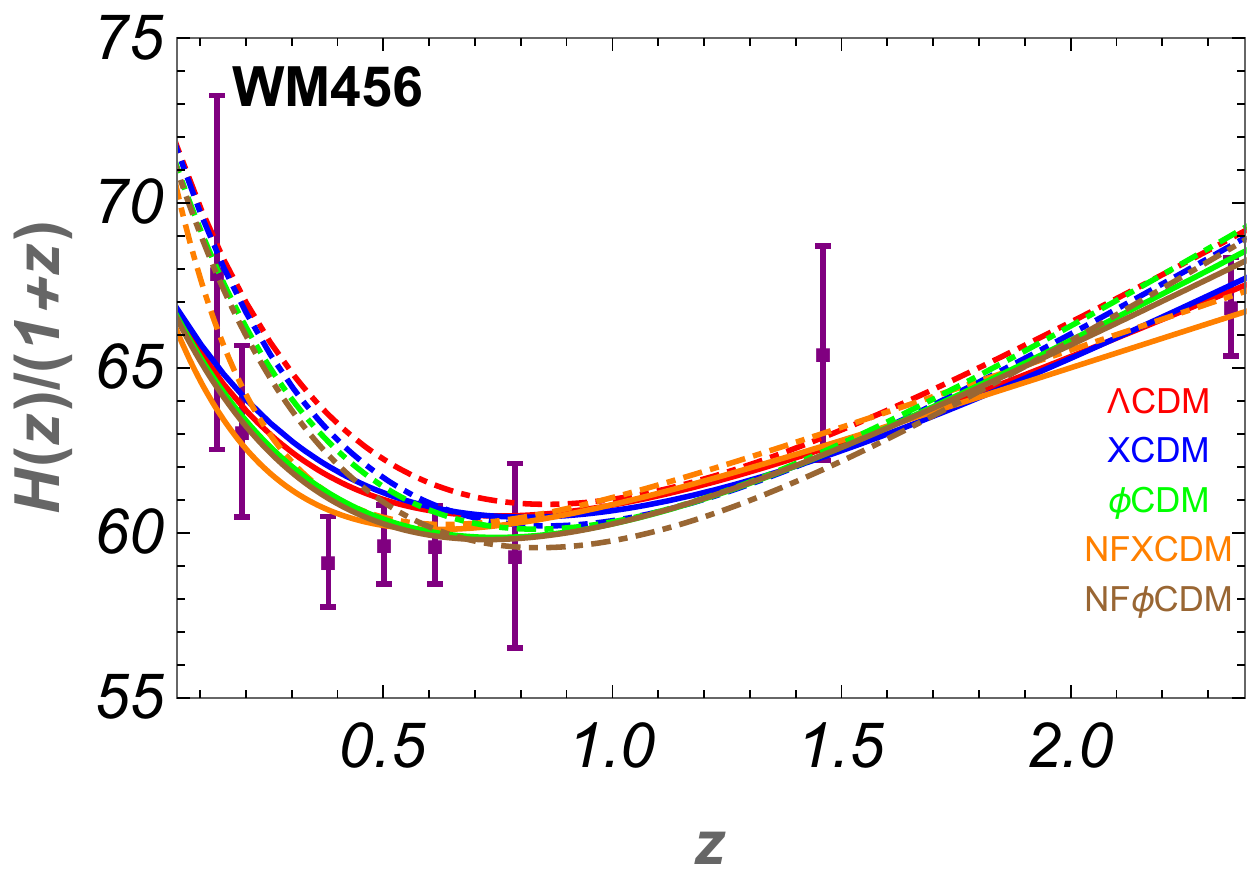}
\caption{
The top panel shows the 38 $H(z)$ measurements of Table \ref{table:Hzdata}.
All error bars are 1$\sigma$. The left (right) panel in the second row shows 
the binned $H(z)$ data with 3 or 4 (4 or 5) measurements per bin, combined 
using weighted mean statistics, listed in Table \ref{table:WM}. In the 
last row, the left (right) panel shows binned $H(z)$ data with 5 or 7 (4, 5, or 6) 
measurements per bin, combined using weighted mean statistics, listed in 
Table \ref{table:WM}. In all panels, there are five different colored solid 
(dot-dashed) best-fit model prediction lines for the two $H_0$ priors used 
in our analyses (see main text for details; NF stands for non-flat).
} \label{fig:Error Plots}
\end{figure}

 In Table \ref{table:Hzdata} we collect 38 Hubble parameter $H(z)$ measurements 
from \cite{Simonetal2005}, \cite{Sternetal2010}, \cite{Morescoetal2012}, 
\cite{Blakeetal2012}, \cite{Zhang2012}, \cite{FontRiberaetal2014}, 
\cite{Delubacetal2015}, \cite{Moresco2015}, \cite{Morescoetal2016}, and 
\cite{Alam2016}. These data are plotted in the top panel of Fig.\ \ref{fig:Error Plots}.

These 38 $H(z)$ measurements are not completely independent. The three 
measurements taken from \cite{Blakeetal2012} are correlated with each other 
and the three measurements of \cite{Alam2016} also are correlated. \textbf{Also, in these and other cases, when BAO observations are used to measure $H(z)$, one has to apply a prior on the radius of the sound horizon, $r_d=\int_{z_d}^{\infty}c_s(z)dz/H(z)$, evaluated at the drag epoch $z_d$, shortly after recombination, when photons and baryons decouple. This prior value of $r_d$ is generally derived from CMB observations.}


Table \ref{table:Hzdata} here is based on Table 1 of \cite{FarooqRatra2013b} 
with the following modifications. We drop older SDSS galaxy clustering $H(z)$ 
determinations from \cite{ChuangWang2013} in favor of the more recent measurements from \cite{Alam2016}. We have added the new 
\cite{Morescoetal2016} measurements. We have dropped the older \cite{Buscaetal2013} Ly-$\alpha$ forest measurement in favor of the newer \cite{FontRiberaetal2014} and \cite{Delubacetal2015} ones. We have also added two new measurements from \cite{Moresco2015}.

There are many other compilations of $H(z)$ data available in the 
literature \citep[see, e.g.,][]{Mengetal2015, Caietal2015, Solaetal2016,
Yuetal2016, Duanetal2016, Qietal2016, Nunesetal2016, ZhangXia2016}. We 
emphasize that our compilation here does not include older, less reliable, 
data, a few with a lot of weight because of anomalously small error bars.

\section{Binning of Hubble parameter data}

There are two reasons to compute ``average" $H(z)$ values for bins in 
redshift space. First, the weighted mean technique of binning data can 
indicate if the original unbinned data have error bars inconsistent with 
Gaussianity, an important consistency check. Second, data binned in 
redshift space can more clearly visually illustrate trends as a function 
of redshift, with the additional advantage of not having to assume a 
particular cosmological model.

The 38 Hubble parameter measurements in Table \ref{table:Hzdata} 
are binned to ensure as many measurements as possible per bin, while also retaining as many (narrow) redshift bins as 

possible. The ideal case is $\sqrt{38}$ measurements in each of 
$\sqrt{38}$ bins. Here we consider about 3-4, 4-5, 4-5-6, and 5-7 measurements 
per bin. The last four measurements are binned by twos in all but 
the 4-5-6 measurement per bin case. In all cases, data points in a given 
bin are not correlated with each other.

After binning the data, we use weighted mean statistics\footnote{We also used median statistics to find central estimates, where the median 
	is the value for which there is a $50\%$ chance of finding a measurement 
	above and below it. Since median statistics does not make use of individual 
	measurement errors, the resultant central estimate error is larger than 
	that for weighted mean statistics. For discussions and applications of median statistics, 
	see \cite{Gottetal2001}, \cite{ChenRatra2003}, \cite{Hodgeetal2009}, 
	\cite{CrandallRatra2014}, \cite{Crandalletal2015}, \cite{Dingetal2015},
	\cite{CrandallRatra2015}, and \cite{Zhengetal2016}.	As in \cite{Farooqetal2013b} for the earlier $H(z)$ data tabulated in 
	\cite{FarooqRatra2013b}, all median statistics analyses results look 
	reasonable and, since the weighted mean results are also all reasonable and 
	more constraining, going forward we use only weighted mean results.} to find a representative central estimate for each bin. Following 
\cite{Podariuetal2001b} the weighted mean is given by,

\begin{equation}
\overline{H}(z)=\frac{\sum_{i=1}^{N}H(z_{i})/\sigma_{i}^{2}}{\sum_{i=1}^{N}1/\sigma_{i}^{2}},
\label{eq:wm}
\end{equation}

where $H(z_{i})$ and $\sigma_{i}$ are the Hubble parameter and one standard deviation of $i=1,2,3...N$ measurements in the bin. We also compute the weighted bin redshift using

\begin{equation}
\overline{z}=\frac{\sum_{i=1}^{N}z_{i}/\sigma_{i}^{2}}{\sum_{i=1}^{N}1/\sigma_{i}^{2}}.
\label{eq:wmz}
\end{equation}

The associated weighted error is given by

\begin{equation}
\overline{\sigma}=\left(\sum_{i=1}^{N}1/\sigma_{i}^{2}\right)^{-1/2}.
\label{eq:wm_error}
\end{equation}

A goodness-of-fit, $\chi^{2}$, can be found for each bin where the reduced $\chi^{2}$ is

\begin{equation}
\chi^{2}_{\nu}=\frac{1}{N-1}\sum_{i=1}^{N}\frac{[H(z_{i})-\overline{H}(z)]^{2}}{\sigma_{i}^{2}}.
\label{eq:chi}
\end{equation}

The number of standard deviations that $\chi_{\nu}$ deviates from unity (the expected value) is given by

\begin{equation}
N_{\sigma}=|\chi_{\nu}-1|\sqrt{2(N-1)}.
\label{eq:nsig}
\end{equation}

A large $N_{\sigma}$ can be the result of non-Gaussian measurements, the presence of un-accounted for systematic errors, or correlations between measurements. Table \ref{table:WM} lists the weighted mean results for the binned $H(z)$ measurements.

The last column of Table \ref{table:WM} shows reasonably small $N_{\sigma}$ 
for all binnings, and so suggests that the error bars of the $H(z)$ data of 
Table \ref{table:Hzdata} are not inconsistent with Gaussianity. As in 
\cite{Farooqetal2013b}, we find that the cosmological constraints that follow 
from the weighted mean binned data are almost identical to those derived using the unbinned data, while the median statistics binned data typically result in somewhat weaker constraints. A possible reason for this could be that some of the unbinned $H(z)$ data error bars might be a bit larger than they really should be. This would be consistent with the low reduced $\chi^{2}$ shown in the last line of Table 1 in \cite{Chenetal2016a}.

The binned data are plotted in the four lower panels of Fig.\ 
\ref{fig:Error Plots}. It is reassuring that, independent of the
binning used, all the binned data sets show clear visual qualitative
evidence for the cosmological deceleration-acceleration transition, as in 
\cite{Farooqetal2013b}. This is model-independent qualitative evidence
for the existence of the cosmological deceleration-acceleration transition. 
We shall see, in Sec.\ 5, that all cosmological models we use in the analysis of
the $H(z)$ data to measure $z_{\rm da}$ result in $z_{\rm da}$ values that 
overlap within the error bars (for a given $H_0$ prior). This is additional
model-independent evidence for the presence of the deceleration-acceleration
transition.


\begin{table*}[!htbp]
	\small
	\caption{Weighted Mean Results For 38 Redshift Measurements}
	\begin{center}
		\begin{threeparttable}
			\centering
			\begin{tabular*}{\textwidth}{l@{\extracolsep{\fill}}cccccc}
				\hline\hline
				\noalign{\vskip 1mm} 

				{\multirow{2}{*}{Bin}}& {\multirow{2}{*}{$N$}}& {\multirow{2}{*}{$z\tnote{a}$}}& {$H(z)$}&{$H(z)$ (1$\sigma$ range)} &{$H(z)$ (2$\sigma$ range)}&{\multirow{2}{*}{$N_\sigma$}}\\

				& & & (km s$^{-1}$ Mpc $^{-1}$)& (km s$^{-1}$ Mpc $^{-1}$)& (km s$^{-1}$ Mpc $^{-1}$)&\\

				\noalign{\vskip 1mm}

				\hline

				\noalign{\vskip 1mm}

				\multicolumn{7}{c}{3 or 4 measurements per bin}\\

				\noalign{\vskip 1mm}

				\hline\\*[-4pt]

				1&	3&	0.0892&	69.0&	59.4-78.5&		49.9-88.0&		2.0\\

				2&	4&	0.185&	76.0&	73.1-78.9&		70.2-81.8&		1.1\\

				3&	3&	0.309&	80.6&	71.0-90.2&		61.5-99.7&		1.5\\

				4&	4&	0.381&	81.5&	79.7-83.4&		77.9-85.2&		1.2\\

				5&	3&	0.438&	85.8&	80.1-91.5&		74.3-97.3&		1.0\\

				6&	3&	0.509&	90.0&	88.1-91.9&		86.3-93.7&		0.53\\

				7&	3&	0.609&	96.5&	94.5-98.4&		92.6-100&			0.22\\

				8&	3&	0.720&	96.6&	91.8-101&			87.0-106&			0.71\\

				9&	4&	0.929&	129&		118-140&			108-151&			0.066\\

				10&	4&	1.43&	158&		149-167&			140-176&			0.047\\

				11&	2&	1.83&	196&		165-227&			133-259&			1.1\\

				12&	2&	2.35&	224&		219-229&			213-234&			0.88\\ [2pt]

				\hline

				\noalign{\vskip 1mm}

				\multicolumn{7}{c}{4 or 5 measurements per bin}\\

				\noalign{\vskip 1mm}

				\hline\\*[-4pt]

				1&	2&	0.0846&	69.0&	58.8-79.2&		48.5-89.5&		1.4\\

				2&	5&	0.184&	75.9&	73.1-78.8&		70.2-81.7&		1.4\\

				3&	5&	0.377&	81.5&	79.7-83.3&		77.8-85.2&		2.3\\

				4&	5&	0.427&	84.6&	79.8-89.4&		75.0-94.2&		1.1\\

				5&	5&	0.518&	90.1&	88.3-91.8&		86.6-93.6&		0.66\\

				6&	4&	0.628&	97.2&	95.3-99.1&		93.3-101&			1.2\\

				7&	4&	0.929&	129&		118-140&			108-151&			0.066\\

				8&	4&	1.43&	158&		149-167&			140-176&			0.047\\

				9&	2&	1.83&	196&		165-227&			133-259&			1.1\\

				10&	2&	2.35&	224&		219-229&			213-234&			0.88\\ [2pt]

				\hline

				\noalign{\vskip 1mm}

				\multicolumn{7}{c}{4, 5, or 6 measurements per bin}\\

				\noalign{\vskip 1mm}

				\hline\\*[-4pt]

				1&	4&	0.137&	77.2&	71.1-83.3&		64.9-89.5&		0.85\\

				2&	5&	0.192&	75.2&	72.1-78.2&		69.1-81.2&		2.3\\

				3&	5&	0.380&	81.6&	79.7-83.4&		77.9-85.2&		1.5\\

				4&	6&	0.502&	89.6&	87.8-91.4&		86.1-93.1&		1.1\\

				5&	4&	0.613&	96.2&	94.3-98.1&		92.4-100&		0.10\\

				6&	6&	0.787&	106&		101-112&	95.8-117&		1.1\\

				7&	6&	1.46&	161&		153-170&		144-178&	0.16\\

				8&	2&	2.35&	224&		219-229&		213-234&		0.88\\[2pt]

				\hline

				\noalign{\vskip 1mm}

				\multicolumn{7}{c}{5 or 7 measurements per bin}\\

				\noalign{\vskip 1mm}

				\hline\\*[-4pt]

				1&	5&	0.166&	75.7&	72.3-79.0&		69.0-82.4&		1.2\\

				2&	7&	0.355&	80.7&	79.0-82.4&		77.2-84.2&		1.6\\

				3&	5&	0.427&	84.6&	79.8-89.4&		75.0-94.2&		1.1\\

				4&	5&	0.518&	90.1&	88.3-91.8&		86.6-93.6&		0.66\\

				5&	7&	0.633&	97.7&	95.8-99.6&		93.9-102&			0.55\\

				6&	5&	1.37&	158&		149-166&			141-174&			0.32\\

				7&	2&	1.83&	196&		165-227&			133-259&			1.1\\

				8&	2&	2.35&	224&		219-229&			213-234&			0.88\\[2pt]

				\hline\hline

			\end{tabular*}

			\begin{tablenotes}

				\item[a] Weighted mean of $z$ values of measurements in the bin.

			\end{tablenotes}

		\end{threeparttable}

	\end{center} 

	\label{table:WM}

\end{table*}




\section{Cosmological Models}

In this section we briefly describe the five models we use to analyze the 
$H(z)$ data. These are the $\Lambda$CDM model that allows for spatial 
curvature and where dark energy is the cosmological constant $\Lambda$ 
\citep[][]{Peebles1984}, as well as the $\phi$CDM model in which dynamical 
dark energy is represented by a slowly evolving scalar field $\phi$ 
\citep[][]{PeeblesRatra1988, RatraPeebles1988}. We also consider an 
incomplete, but popular, parameterization of dynamical dark energy, XCDM, 
where dynamical dark energy is represented by an $X$-fluid. In the 
$\phi$CDM and XCDM cases, we consider both spatially-flat and non-flat models
\citep[][]{Pavlovetal2013}.

In the $\Lambda$CDM model with spatial curvature the Hubble parameter is
\begin{eqnarray}
H(z;H_0,\textbf{p})&=& H_0 \left[\Omega_{m0} (1+z)^3 + \Omega_{\Lambda}+(1-\Omega_{m0}-\Omega_\Lambda) (1+z)^2 \right]^{1/2},
\label{eq:LCDMFM}
\end{eqnarray}
where we have made use of $\Omega_{K0} = 1-\Omega_{m0}-\Omega_{\Lambda}$ to 
eliminate the current value of the space curvature energy density parameter 
in favor of the current value of the non-relativistic matter energy density 
parameter, $\Omega_{m0}$, and the cosmological constant energy density parameter, $\Omega_{\Lambda}$. Here $\textbf{p}=(\Omega_{m0},\Omega_{\Lambda})$
are the two cosmological parameters that conventionally characterize 
$\Lambda$CDM and $H_0$ is the value of Hubble parameter at the present time 
and is called the Hubble constant.

It has become fashionable to parameterize dynamical dark energy as a 
spatially homogeneous $X$-fluid, with a constant equation of state 
parameter, $\omega_{X}=p_{X}/\rho_{X} <-1/3 $ (here $p_{X}$ and 
$\rho_{X}$ are the pressure and energy density of the $X$-fluid 
respectively). For the spatially-flat XCDM parameterization, using 
$\Omega_{X0}=1-\Omega_{m0}$ (where $\Omega_{X0}$ is the current value of 
the $X$-fluid energy density parameter), we have
\begin{eqnarray}
H(z; H_0, \textbf{p})&=&H_0 [\Omega_{m0}(1+z)^3 + (1 - \Omega_{m0}) (1+z)^{3(1+\omega_{X})}]^{1/2}.
\label{eq:FXCDMFM}   
\end{eqnarray}

In this spatially-flat case the two cosmological parameters are 
$\textbf{p}=(\Omega_{m0},\omega_{X})$. The XCDM parameterization is 
incomplete as it cannot describe the evolution of energy density 
inhomogeneities. In the non-flat XCDM parametrization case, $\Omega_{K0}$ 
is the third free parameter and
\begin{eqnarray}
H(z; H_0, \textbf{p})&=&H_0 [\Omega_{m0}(1+z)^3 + (1-\Omega_{m0}-\Omega_{K0})(1+z)^{3(1+\omega_{X})}+\Omega_{K0}(1+z)^2\ ]^{1/2},
\label{eq:FXCDMFM1}   
\end{eqnarray}
where the three cosmological parameters are $\textbf{p}=({\Omega_{m0},\omega_X, \Omega_{K0}})$. $\phi$CDM is the simplest, complete and consistent dynamical dark energy model. Here dark energy is modeled as a slowly-rolling scalar field $\phi$ with an, e.g., inverse-power-law potential energy density $V(\phi)=\kappa m_p^2 \phi^{-\alpha}/2 $, where $m_{p}$ is the Planck mass and $\alpha$ is a non-negative parameter that determines the coefficient $\kappa$($m_{p}$,$\alpha$) \citep[][]{PeeblesRatra1988}. The equation of motion of the scalar field is 
\begin{eqnarray}
\ddot{\phi} + 3 \frac{\dot{a}}{a}\dot{\phi}-\frac{\kappa}{2} \alpha m_p^2 \phi^{-(\alpha+1)} = 0,
\label{eq:Equationofmotionphi} 
\end{eqnarray}
where an overdot represents a time derivative and $a$ is the scale factor. For the spatially-flat $\phi$CDM model
\begin{eqnarray}
H(z; H_0, \textbf{p})&=&H_0[\Omega_{m0}(1+z)^3+\Omega_\phi (z,\alpha)]^{1/2},
\label{eq:FphiCDMFM}
\end{eqnarray}
where the time-dependent scalar field energy density parameter is
\begin{eqnarray}
\Omega_\phi(z,\alpha)=\frac{1}{12H^{2}_{0}}\left(\dot\phi^2+\kappa m^{2}_{p}\phi^{-\alpha}\right).
\label{eq:Omegaphi} 
\end{eqnarray}

In this case the two cosmological parameters are $\textbf{p}=({\Omega_{m0},\alpha})$. In the non-flat $\phi$CDM model
\begin{eqnarray}
H(z; H_0, \textbf{p})&=&H_0[\Omega_{m0}(1+z)^3+\Omega_\phi (z,\alpha)+ \Omega_{K0}(1+z)^2]^{1/2},
\label{eq:NFphiCDMFM}
\end{eqnarray}
and the three cosmological parameters are $\textbf{p}=({\Omega_{m0},\alpha, \Omega_{K0}})$.

Solving the coupled differential equations of motion allows for a numerical computation of the Hubble parameter $H(z; H_0, \textbf{p})$ \citep[][]{PeeblesRatra1988, Samushia2009, Farooq2013, Pavlovetal2013}.\footnote{For discussions of observational constraints on the $\phi$CDM model see, e.g. \cite{PodariuRatra2000}, \cite{ChenRatra2004}, \cite{SamushiaRatra2010}, \cite{Samushiaetal2010}, \cite{Campanellietal2012}, \cite{Pavlovetal2014}, \cite{Avsajanishvilietal2014}, \cite{Avsajanishvilietal2015}, \cite{Limaetal2016}, \cite{GosencaColes2015}, and \cite{Chenetal2016b}.}

In Sec.\ \ref{Constraints} we use these expressions for the Hubble parameter 
in conjunction with the $H(z)$ measurements in Table \ref{table:Hzdata} to 
constrain the cosmological parameters of these models. In our analyses here we
study the following parameter ranges: $0 \leq \Omega_{m0} \leq 1$, 
$0 \leq \Omega_{\Lambda} \leq 1.4$, $-2 \leq \omega_{X} \leq 0$, 
$0 \leq \alpha \leq 5$, and $-0.7 \leq \Omega_{K0} \leq 0.7$ for non-flat
XCDM and $-0.4 \leq \Omega_{K0} \leq 0.4$ for non-flat $\phi$CDM \citep[which is double the $\Omega_{K0}$ range used in][]{Farooqetal2015}.

\section{Cosmological deceleration-acceleration transition redshift}

At the current epoch, dark energy dominates the cosmological energy budget and accelerates the cosmological expansion. At earlier times non-relativistic (baryonic and cold dark) matter dominated the energy budget and the cosmological expansion decelerated. The cosmological deceleration-acceleration transition redshift, $z_{\mathrm{da}}$, is defined as the redshift at which $\ddot{a}=0$, in the cosmological model under consideration. $\ddot{a}$ is proportional to the active gravitational mass density, the sum of the energy densities and three times the pressure of the constituents.

For $\Lambda$CDM, setting $\ddot{a}=0$ we find
\begin{eqnarray}
z_{\mathrm{da}}&=&\left(\frac{2\Omega_\Lambda}{\Omega_{m0}}\right)^{1/3}-1.
\label{eq:LCDMzda3}
\end{eqnarray}

For the case of the spatially-flat XCDM parameterization
\begin{eqnarray}
z_{\mathrm{da}}&=& \left(\frac{\Omega_{m0}}{\left(\Omega_{m0}-1\right)\left(1+3\omega_X\right)}\right)^{1/{3\omega_X}}-1,
\label{eq:FXCDMzda2}
\end{eqnarray}
while for non-flat XCDM
\begin{eqnarray}
z_{\mathrm{da}}&=& \left(\frac{\Omega_{m0}}{\left(\Omega_{m0}+\Omega_{K0}-1\right)\left(1+3\omega_X\right)}\right)^{1/{3\omega_X}}-1.
\label{eq:NFXCDMzda1}
\end{eqnarray}
For the spatially-flat $\phi$CDM model, defining the time-dependent equation-of-state-parameter for the scalar field
\begin{eqnarray}
\omega_\phi(z)=\frac{\frac{1}{2}\dot{\phi}^2-V(\phi)}{\frac{1}{2}\dot{\phi}^2+V(\phi)},
\label{eq:FPCDMzda4}
\end{eqnarray}
the redshift $z_{\rm da}$ ($\Omega_{m0}$, $\alpha$) is determined by numerically solving
\begin{eqnarray}
\Omega_{m0} (1+z_{\rm da})^3+\Omega_{\phi}(z_{\mathrm{da}}, \alpha)\left[1+3~\omega_{\phi}(z_{\rm da})\right]=0
\label{eq:FPCDMzda5}
\end{eqnarray}
where $\Omega_{\phi 0} = 1 - \Omega_{m0}$. In the non-flat $\phi$CDM model $z_{\rm da}$ ($\Omega_{m0}$, $\alpha$, $\Omega_{K0}$) is determined by numerically solving the same equation, but now setting $\Omega_{\phi 0} = 1 - \Omega_{m0} - \Omega_{K0}$.

\begin{center}
\begin{threeparttable}
\caption{Deceleration-Acceleration Transition Redshifts\tnote{a}} 
\vspace{5 mm}
\begin{tabular}{cccccc}
\hline\hline
\multirow{2}{*}{Model} &\multirow{2}{*}{$h$ Prior\tnote{b}} & \multirow{2}{*}{BF\tnote{c}} & \multirow{2}{*}{$\chi_{\rm min}^2$} & \multirow{2}{*}{$z_{\rm da} \pm \sigma_{z_{\rm da}}$\tnote{d}} & \multirow{2}{*}{$z_{\rm da} \pm \sigma_{z_{\rm da}}$\tnote{e}}\\
{}&{}&{}&{}&{}& {} \\
\hline\\*[-6pt]

\multirow{4}{*}{$\Lambda$CDM} & \multirow{2}{*}{0.68  $\pm$ 0.028} & $\Omega_{m0}=0.23$   &	\multirow{2}{*}{22.4}  &	\multirow{2}{*}{0.723 $\pm$ 0.089} &	   \multirow{2}{*}{0.690 $\pm$ 0.096}\\

{}&{}&{$\Omega_{\Lambda}=0.60$}&{}& {} & {}\\

\cline{2-6}\\*[-6pt]

{}& \multirow{2}{*}{$0.7324 \pm 0.0174$}& $\Omega_{m0}=0.25$   &	\multirow{2}{*}{24.2}   &	\multirow{2}{*}{0.832 $\pm$ 0.055} &	   \multirow{2}{*}{0.781 $\pm$ 0.067}\\

{}&{}&{$\Omega_{\Lambda}=0.78$}&{}&{}&{} \\

\hline\\*[-4pt]

\multirow{4}{*}{Flat XCDM} & \multirow{2}{*}{0.68  $\pm$ 0.028}  &	$\Omega_{m0}=0.26$  &	\multirow{2}{*}{22.5} &	\multirow{2}{*}{0.753 $\pm$ 0.091}&	   \multirow{2}{*}{0.677 $\pm$ 0.097}\\

{}&{}&{$\omega_{X}=-0.86$}&{}&{}& {} \\

\cline{2-6}\\*[-6pt]

{}& \multirow{2}{*}{$0.7324 \pm 0.0174$}  &	$\Omega_{m0}=0.24$  &	\multirow{2}{*}{23.9}  &	\multirow{2}{*}{0.813 $\pm$ 0.062} &	   \multirow{2}{*}{0.696 $\pm$ 0.082}\\

{}&{}&{$\omega_{X}=-1.06$}&{}& {}& {} \\

\hline\\*[-6pt]

\multirow{4}{*}{Flat $\phi$CDM} & \multirow{2}{*}{0.68  $\pm$ 0.028} &	 $\Omega_{m0} =0.27$ &	\multirow{2}{*}{22.9}  &	\multirow{2}{*}{0.703 $\pm$ 0.104}&	   \multirow{2}{*}{0.724 $\pm$ 0.148} \\

{}&{}&{$\alpha=0.50$}&{}& {}& {}  \\

\cline{2-6}\\*[-6pt]

{}& \multirow{2}{*}{$0.7324 \pm 0.0174$}&	 $\Omega_{m0}=0.25$&	\multirow{2}{*}{25.2} &	\multirow{2}{*}{$0.885 \pm 0.056$}  &	   \multirow{2}{*}{0.850 $\pm$ 0.116}\\

{}&{}&{$\alpha=0$}&{}& {} & {}\\

\hline\\*[-4pt]

\multirow{6}{*}{Non-flat XCDM} & \multirow{3}{*}{0.68 $\pm$ 0.028} &	$\Omega_{m0}=0.15$&	\multirow{3}{*}{21.9}  &	\multirow{3}{*}{0.684 $\pm$ 0.117} &	   \multirow{3}{*}{$\cdots$}\\

{}&{}&{$\omega_{X}=-1.68$}&{}& {}& {} \\

{}&{}&{$\Omega_{K0}=0.45$}&{}& {}& {} \\

\cline{2-6}\\*[-6pt]

{}& \multirow{3}{*}{$0.7324 \pm 0.0174$}&	$\Omega_{m0}=0.13$&	\multirow{3}{*}{20.3}  &	\multirow{3}{*}{0.709 $\pm$ 0.090} &	   \multirow{3}{*}{$\cdots$}\\

{}&{}&{$\omega_{X}=-2$}&{}&{}&  {} \\

{}&{}&{$\Omega_{K0}=0.41$}&{}&{}&  {} \\

\hline\\*[-6pt]

\multirow{6}{*}{Non-flat $\phi$CDM} & \multirow{3}{*}{0.68 $\pm$ 0.028}&	$\Omega_{m0}=0.23$ &	\multirow{3}{*}{22.6}  &	\multirow{3}{*}{0.690 $\pm$ 0.118} &	   \multirow{3}{*}{$\cdots$}\\

{}&{}&{$\alpha=0$}&{}&{}&  {} \\

{}&{}&{$\Omega_{K0}=0.18$}&{}&{}&  {} \\

\cline{2-6}\\*[-6pt]

{}& \multirow{3}{*}{$0.7324 \pm 0.0174$}&	$\Omega_{m0}=0.25$&	\multirow{3}{*}{25.0}  &	\multirow{3}{*}{0.853 $\pm$ 0.053} &	   \multirow{3}{*}{$\cdots$}\\

{}&{}&{$\alpha=0$}&{}&{}&  {}  \\

{}&{}&{$\Omega_{K0}=-0.03$}&{}&{}&  {}  \\

\hline

\hline

\end{tabular}

\begin{tablenotes}

\item[a]{Estimated using the unbinned data of Table \ref{table:Hzdata}.}
\item[b]{Hubble constant in units of 100 km s$^{-1}$ Mpc$^{-1}$.}
\item[c]{Best-fit parameter values.}
\item[d]{Computed using Eqs.\ \ref{eq:LCDMzda3}---\ref{eq:Expectedzda1} of this work.}
\item[e]{The deceleration-acceleration transition redshift in the model, as computed in \cite{Farooqetal2013b} Table 1. Note that the best-fit cosmological parameter values found in \cite{Farooqetal2013b} differ from those found here and listed in this Table.}
\end{tablenotes}
\label{table:Un-binned data zda}
\end{threeparttable}
\end{center}

To compute the expected values $\langle z_{\rm da} \rangle$ and $\langle z_{\rm da}^2 \rangle$ for the two-parameter models we use
\begin{eqnarray}
\langle z_{\mathrm{da}} \rangle=\frac{\int\int z_{\rm da}(\textbf{p}) \mathcal{L}(\textbf{p}) d\textbf{p}}{\int\int \mathcal{L}(\textbf{p}) d\textbf{p}},~~~~~~~~~~~~
\langle z_{\rm da}^2 \rangle=\frac{\int\int z_{\rm da}^2(\textbf{p}) \mathcal{L}(\textbf{p}) d\textbf{p}}{\int\int \mathcal{L}(\textbf{p}) d\textbf{p}}.
\label{eq:Expectedzda1}
\end{eqnarray} 

Here $\mathcal{L}(\textbf{p})$ is the $H(z)$ data likelihood function after marginalization over the Gaussian $H_0$ prior in the two-parameter model 
under consideration, as explained in \cite{Farooqetal2013a} and 
\cite{Farooqetal2015} but this time accounting for the non-diagonal 
correlation matrices of the \cite{Blakeetal2012} and the \cite{Alam2016} measurements, which have a small effect. $\mathcal{L}(\textbf{p})$
depends only on the model parameters 
$(\Omega_{m0},\Omega_{\Lambda})$ for $\Lambda$CDM, $(\Omega_{m0},\omega_{X})$ 
for flat XCDM, and $(\Omega_{m0},\alpha)$ for flat $\phi$CDM. The 
generalization for the three-parameter models is straightforward. The 
standard deviation in $z_{\rm da}$ is computed from the standard formula 
$\sigma_{z_{\rm da}}=\sqrt{\langle z_{\rm da}^2 \rangle-\langle z_{\rm da} \rangle^2}$. The results of this computation are summarized in 
Table \ref{table:Un-binned data zda}.

Table \ref{table:Un-binned data zda} shows best-fit cosmological parameter
values and the corresponding minimum $\chi^2$ for the five different 
cosmological models and for the two Gaussian $H_0$ priors. The second last 
column in Table \ref{table:Un-binned data zda} shows the average deceleration-acceleration transition 
redshift with corresponding standard deviation for each model. It is very 
reassuring that the $z_{\rm da}$ values we measure in the five different 
models (for a given $H_0$ prior) overlap reasonably well. (The main effect
on the measured $z_{\rm da}$ value is the assumed $H_0$ prior value.) 
Given that the measured $z_{\rm da}$ are almost independent of the other 
model parameters, within the errors, we may conclude that to leading order 
we have measured a model-independent $z_{\rm da}$ value. However, it is useful 
to have a single summary value for this cosmological parameter.

By taking the simple average of the penultimate column $z_{\rm da}$ values 
and computing the population standard deviation for the five values in this column, we find $z_{\rm da}=0.71 \pm 0.03$ ($0.82 \pm 0.06$) for $H_0 \pm \sigma_{H_0}= 68 \pm 2.8$ ($73.24 \pm 1.74$) km s$^{-1}$ Mpc$^{-1}$. Using all ten $z_{\rm da}$ values in the penultimate column of Table \ref{table:Un-binned data zda} we find $z_{\rm da}=0.76 \pm 0.07$.

A more reliable summary value of the deceleration-acceleration transition 
redshift is determined from a weighted mean analysis. Using Eqs.\ 
\ref{eq:wm}---\ref{eq:wm_error}, we find $z_{\rm da}=0.72 \pm 0.05$ 
($0.84 \pm 0.03$) for $H_0 \pm \sigma_{H_0}=68 \pm 2.8$ ($73.24 \pm 1.74$) 
km s$^{-1}$ Mpc$^{-1}$, and using all ten values in the penultimate column 
of Table \ref{table:Un-binned data zda} we get $z_{\rm da}=0.80 \pm 0.02$. \textbf{By looking at the fourth and the fifth columns of Table\ \ref{table:Un-binned data zda} it appears that all the five models discussed here fit better with the lower value of $H_0$ while the uncertainty in $z_{da}$ is more sensitive to $\sigma_{H_0}$.}

These results are listed in Table \ref{table:zdasummaryvalues} 
and compared with the previously computed summary values of 
\cite{Farooqetal2013b}. Note that only three models ($\Lambda$CDM, flat
XCDM, and flat $\phi$CDM) were considered in \cite{Farooqetal2013b}.
Here we also consider non-flat XCDM and non-flat $\phi$CDM. We see that there
is good agreement between the old and new weighted mean $z_{\rm da}$
for $h = 0.68$, less so for $h = 0.7324$. From Table 
\ref{table:zdasummaryvalues} we see that for a given $H_0$ the weighted 
average values of $z_{\rm da}$ for all five models and for the two sets of (non-nested) triplets of models agree to within the error bars.

\vspace{-10mm}
\begin{center}
\begin{threeparttable}
\caption{$z_{\rm da}$ Summary}
\vspace{5 mm}
\begin{tabular}{ccccccc}
\hline\hline
\multirow{4}{*}{} &\multicolumn{2}{c}{\multirow{2}{*}{$h\pm \sigma_{h}=0.68 \pm 0.028$\tnote{a}}} & \multicolumn{2}{c}{\multirow{2}{*}{$h\pm \sigma_{h}=0.7324 \pm 0.0174$\tnote{a}}} & \multicolumn{2}{c}{\multirow{2}{*}{Total\tnote{b}}} \\
&&&&&&\\
{}&\multirow{2}{*}{Here\tnote{c}}&\multirow{2}{*}{Previous\tnote{d}}&\multirow{2}{*}{Here\tnote{c}}&\multirow{2}{*}{Previous\tnote{d}}&\multirow{2}{*}{Here\tnote{c}}&\multirow{2}{*}{Previous\tnote{d}}\\
&&&&&&\\
\hline
\multirow{2}{*}{Simple Averages}&\multirow{2}{*}{$0.71 \pm 0.03$}&\multirow{2}{*}{$0.70\pm 0.02$}&\multirow{2}{*}{$0.82\pm0.06$}&\multirow{2}{*}{$0.78\pm0.06$}&\multirow{2}{*}{$0.76\pm0.07$}&\multirow{2}{*}{$0.74\pm 0.06$} \\
&&&&&&\\
\hline
\multirow{2}{*}{Weighted Averages}&\multirow{2}{*}{$0.72 \pm 0.05$}&\multirow{2}{*}{$0.69\pm 0.06$}&\multirow{2}{*}{$0.84\pm0.03$}&\multirow{2}{*}{$0.76\pm0.05$}&\multirow{2}{*}{$0.80\pm0.02$}&\multirow{2}{*}{$0.74\pm 0.04$} \\
&&&&&&\\
\hline
\multirow{2}{*}{Simple Averages from}&\multirow{4}{*}{$0.73\pm0.02$}&\multirow{4}{*}{$\cdots$}&\multirow{4}{*}{$0.84\pm 0.03$}&\multirow{4}{*}{$\cdots$}&\multirow{4}{*}{$0.78\pm0.06$}&\multirow{4}{*}{$\cdots$}\\
&&&&&&\\
\multirow{2}{*}{$\Lambda$CDM and Flat Models}&&&&&&\\
&&&&&&\\
\hline
\multirow{2}{*}{Weighted Averages from}&\multirow{4}{*}{$0.73\pm0.05$}&\multirow{4}{*}{$\cdots$}&\multirow{4}{*}{$0.85\pm 0.03$}&\multirow{4}{*}{$\cdots$}&\multirow{4}{*}{$0.81\pm0.03$}&\multirow{4}{*}{$\cdots$}\\
&&&&&&\\
\multirow{2}{*}{$\Lambda$CDM and Flat Models}&&&&&&\\
&&&&&&\\
\hline
\multirow{2}{*}{Simple Averages from}&\multirow{4}{*}{$0.70\pm0.02$}&\multirow{4}{*}{$\cdots$}&\multirow{4}{*}{$0.80\pm 0.06$}&\multirow{4}{*}{$\cdots$}&\multirow{4}{*}{$0.75\pm0.07$}&\multirow{4}{*}{$\cdots$}\\
&&&&&&\\
\multirow{2}{*}{Non-Flat Models}&&&&&&\\
&&&&&&\\
\hline
\multirow{2}{*}{Weighted Averages from}&\multirow{4}{*}{$0.70\pm0.06$}&\multirow{4}{*}{$\cdots$}&\multirow{4}{*}{$0.82\pm 0.04$}&\multirow{4}{*}{$\cdots$}&\multirow{4}{*}{$0.79\pm0.03$}&\multirow{4}{*}{$\cdots$}\\
&&&&&&\\
\multirow{2}{*}{Non-Flat Models}&&&&&&\\
&&&&&&\\
\hline
\hline
\end{tabular}
\begin{tablenotes}
\item[a]{Hubble constant in units of 100 km s$^{-1}$ Mpc$^{-1}$.}
\item[b]{Combination of results from both $H_0$ priors.}
\item[c]{Estimated using the unbinned data of 38 $H(z)$ measurements from Table \ref{table:Hzdata}.}
\item[d]{Results from \cite{Farooqetal2013b}. We have corrected typos in that paper here.}
\end{tablenotes}
\label{table:zdasummaryvalues}
\end{threeparttable}
\end{center}

\section{Cosmological parameter constraints}
\label{Constraints}

In this section, we use the 38 Hubble parameter measurements 
(over $0.07 \leq z \leq 2.36$) listed in Table \ref{table:Hzdata} to 
determine constraints on the parameters of the five different cosmological 
models. We use the technique of \cite{Farooqetal2015} to find constraints 
on $(\Omega_{m0}, \Omega_{\Lambda})$ in the $\Lambda$CDM model, 
$(\Omega_{m0}, \omega_{X})$ for the spatially-flat XCDM parameterization, 
$(\Omega_{m0}, \alpha)$ in the spatially-flat $\phi$CDM model, 
$(\Omega_{m0}, \omega_{X}, \Omega_{K0})$ for the XCDM parameterization 
with space curvature, and $(\Omega_{m0}, \alpha, \Omega_{K0})$ in the 
$\phi$CDM model with space curvature. For the $H(z)$ cosmological test, 
cosmological parameter constraints depend on the value of the Hubble 
constant \citep[see, e.g.,][]{Samushiaetal2007}. We use two different 
Gaussian priors for the Hubble constant; the lower value is $68 \pm 2.8$ 
km s$^{-1}$ Mpc$^{-1}$ and the higher is $73.24 \pm 1.74$ km s$^{-1}$ Mpc$^{-1}$.
The lower value is from a median statistics analysis \citep{Gottetal2001} 
of 553 measurements of $H_0$ tabulated by Huchra \citep{ChenRatra2011a}. 
It agrees with earlier median statistics estimates of $H_{0}$ from smaller
compilations \citep[][]{Gottetal2001, Chenetal2003} and is consistent with a number of other recent determinations 
of $H_0$ from Wilkinson Microwave Anisotropy Probe, Atacama Cosmology 
Telescope, and Planck CMB anisotropy data \citep[][]{Hinshawetal2013, Sieversetal2013, Adeetal2015, Addisonetal2016}, 
from BAO measurements \citep[][]{Aubourgetal2015, Rossetal2015, LHuillierShafieloo2016}, 
from Hubble parameter data \citep[][]{Chenetal2016a}, and with what is 
expected in the standard model of particle physics with only three light 
neutrino species given current cosmological data 
\citep[see, e.g.][]{Calabreseetal2012}. The higher value is a relatively 
local measurement, based on Hubble Space Telescope data 
\citep{Riessetal2016}. It is consistent with 
other recent local measurements of $H_0$ \citep[][]{Riessetal2011, Freedmanetal2012, Efstathiou2014}.

We compute the likelihood function $\mathcal{L}(\textbf{p})$ for the 
models under discussion using Eq.\ (18) of \cite{Farooqetal2013a} 
for the ranges of the cosmological parameters listed at the end of
Sec.\ 4. We need these likelihood functions for the $z_{\rm da}$ 
computation of the previous section, which is the main result of the paper.
In this section, we use these likelihood functions to constrain 
cosmological parameters such as the dark energy density.

\begin{figure}[H]

\centering
  \includegraphics[width=59.5mm]{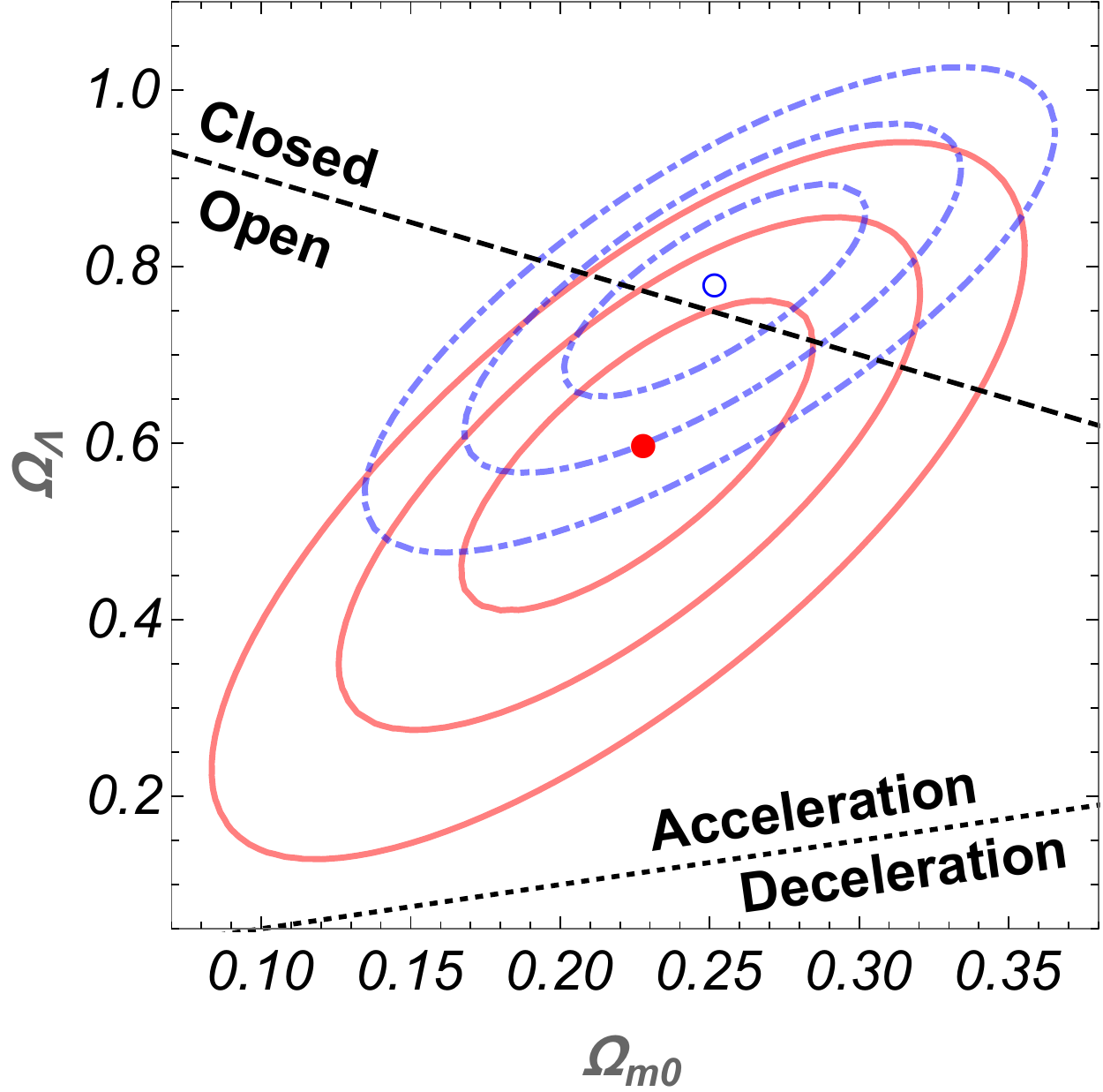}
  \includegraphics[width=59.5mm]{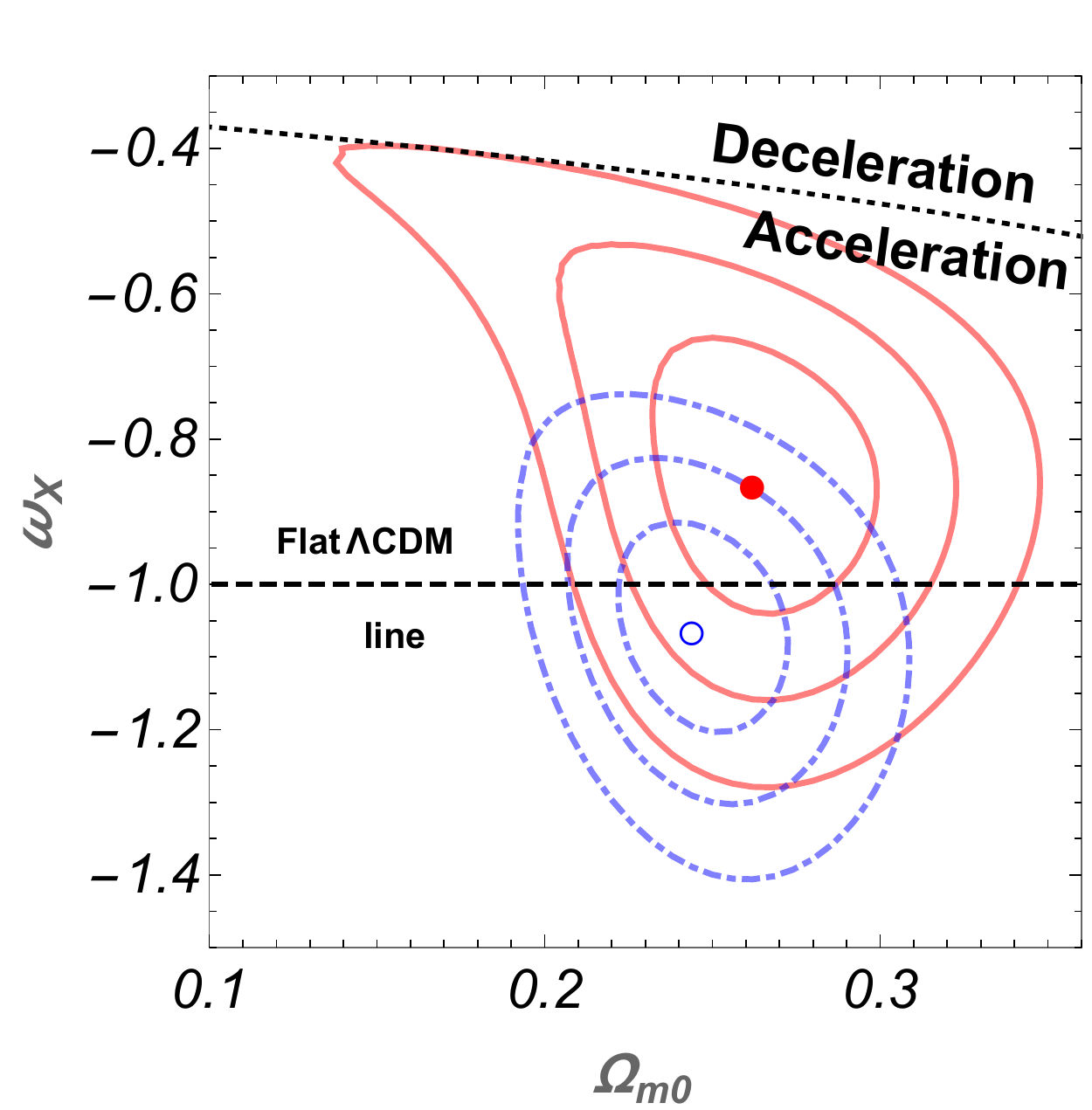}
  \includegraphics[width=57mm]{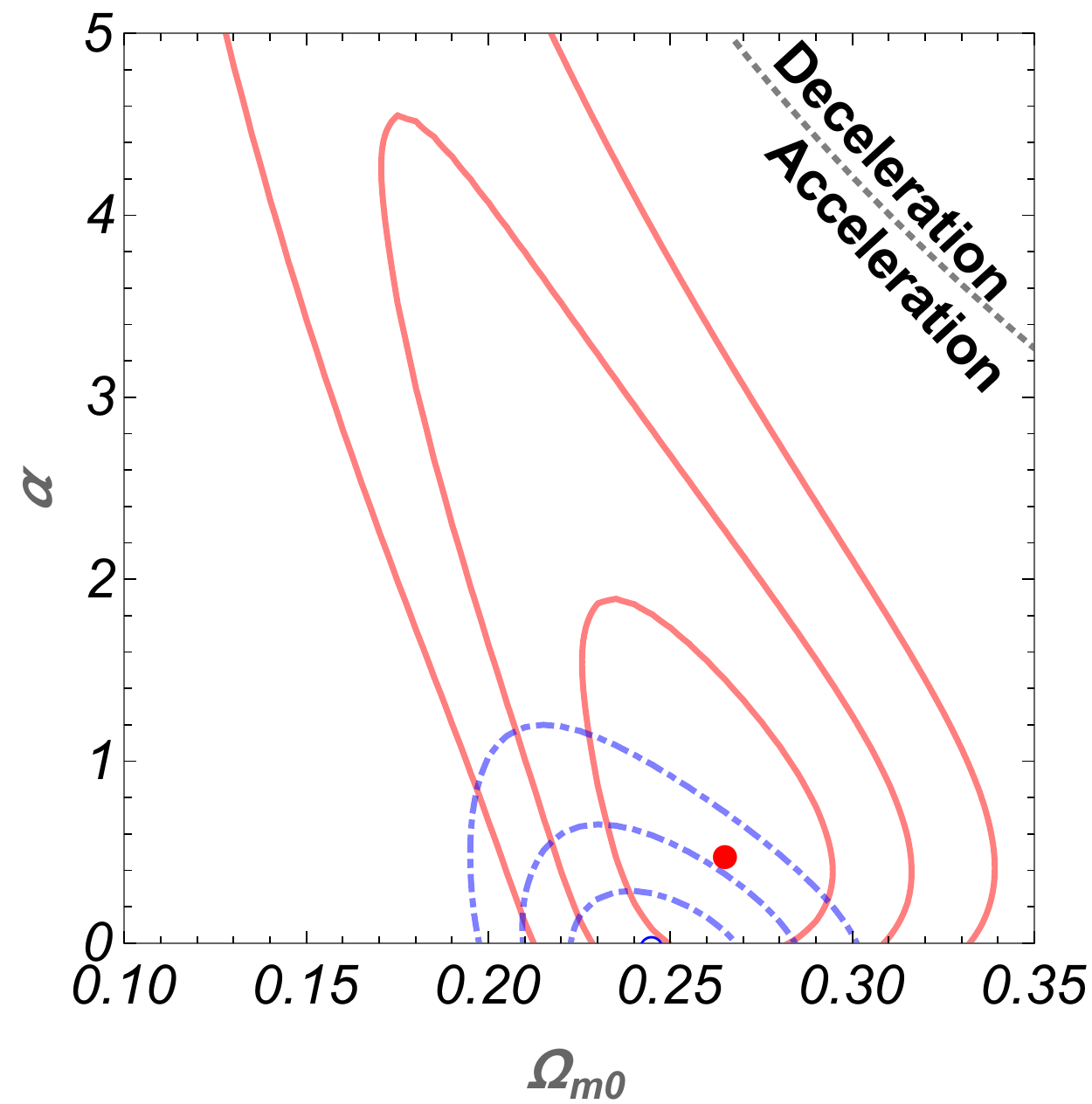}
\caption{
The three panels (from left to right) show $1\sigma$, $2\sigma$, and $3\sigma$ 
red solid (blue dot-dashed) constraint contours for the lower (higher) $H_0$ prior,
for $\Lambda$CDM, flat XCDM, and flat $\phi$CDM respectively. Red solid (blue empty) 
circles are the best-fit points for the lower (higher) $H_0$ prior. The 
straight dashed lines in the first and second panels correspond to 
spatially-flat $\Lambda$CDM models, the dotted lines demarcate 
zero-acceleration models, and the shaded area in the upper left-hand corner of the left panel is the 
region for which there is no big bang. For quantitative parameter best-fit 
values and ranges see Tables \ref{table:Un-binned data zda} and 
\ref{table:1Dfor2parametermodels}.
} \label{fig:Flatmodelsconstraints}
\end{figure}

For the two-parameter models, maximizing the likelihood function 
$\mathcal{L}(\mathbf{p})$ is performed by minimizing the corresponding 
$\chi^2(\mathbf{p})\equiv -2\mathrm{ln}[\mathcal{L}(\mathbf{p})]$ 
following the procedure of \cite{Farooqetal2015}. The corresponding 
minimum values of $\chi^2$ and best-fit parameter values for the 
two-parameter models are summarized in Table \ref{table:Un-binned data zda}.
$1\sigma$, $2\sigma$, and $3\sigma$ confidence contours are computed  
following the procedure of \cite{Farooqetal2015} and results are shown 
in Fig.\ \ref{fig:Flatmodelsconstraints}. The generalization of this procedure 
for the three-parameter models is straightforward and best-fit 
three-dimensional parameter values and minimum $\chi^2$ are also summarized 
in Table \ref{table:Un-binned data zda}.

\begin{figure}[H]
\centering
  \includegraphics[width=59.5mm]{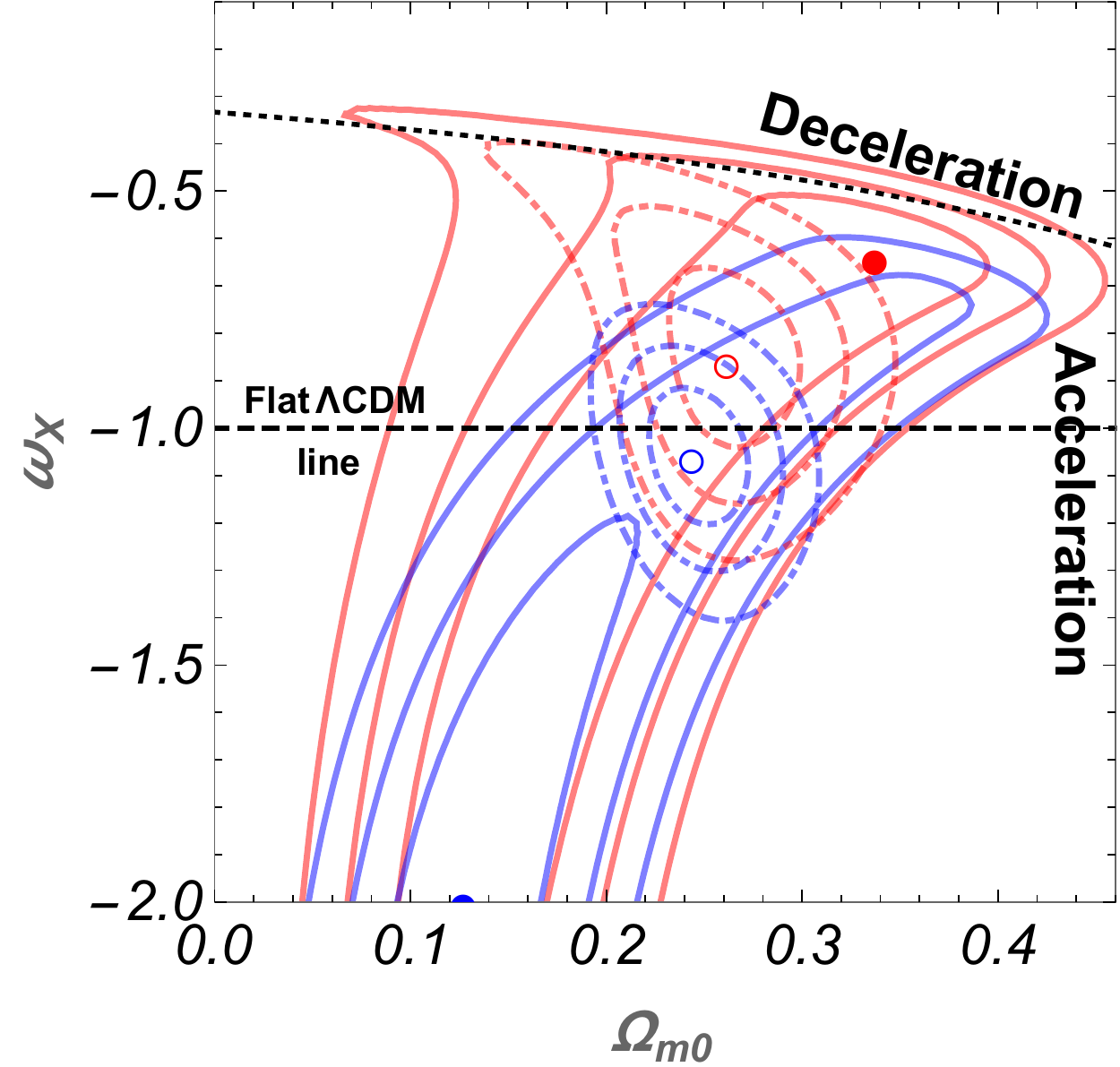}
  \includegraphics[width=59.5mm]{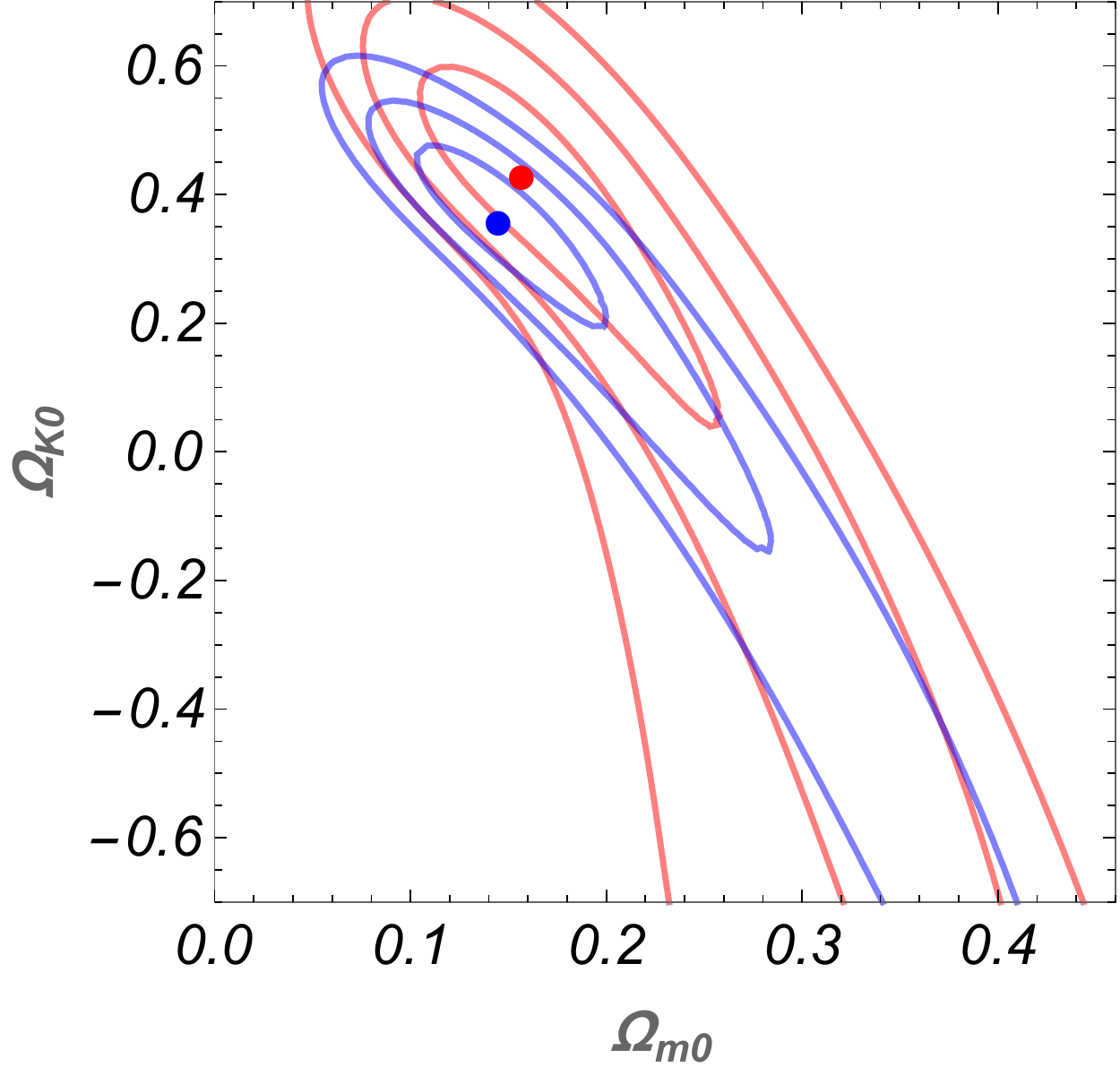}
  \includegraphics[width=59.5mm]{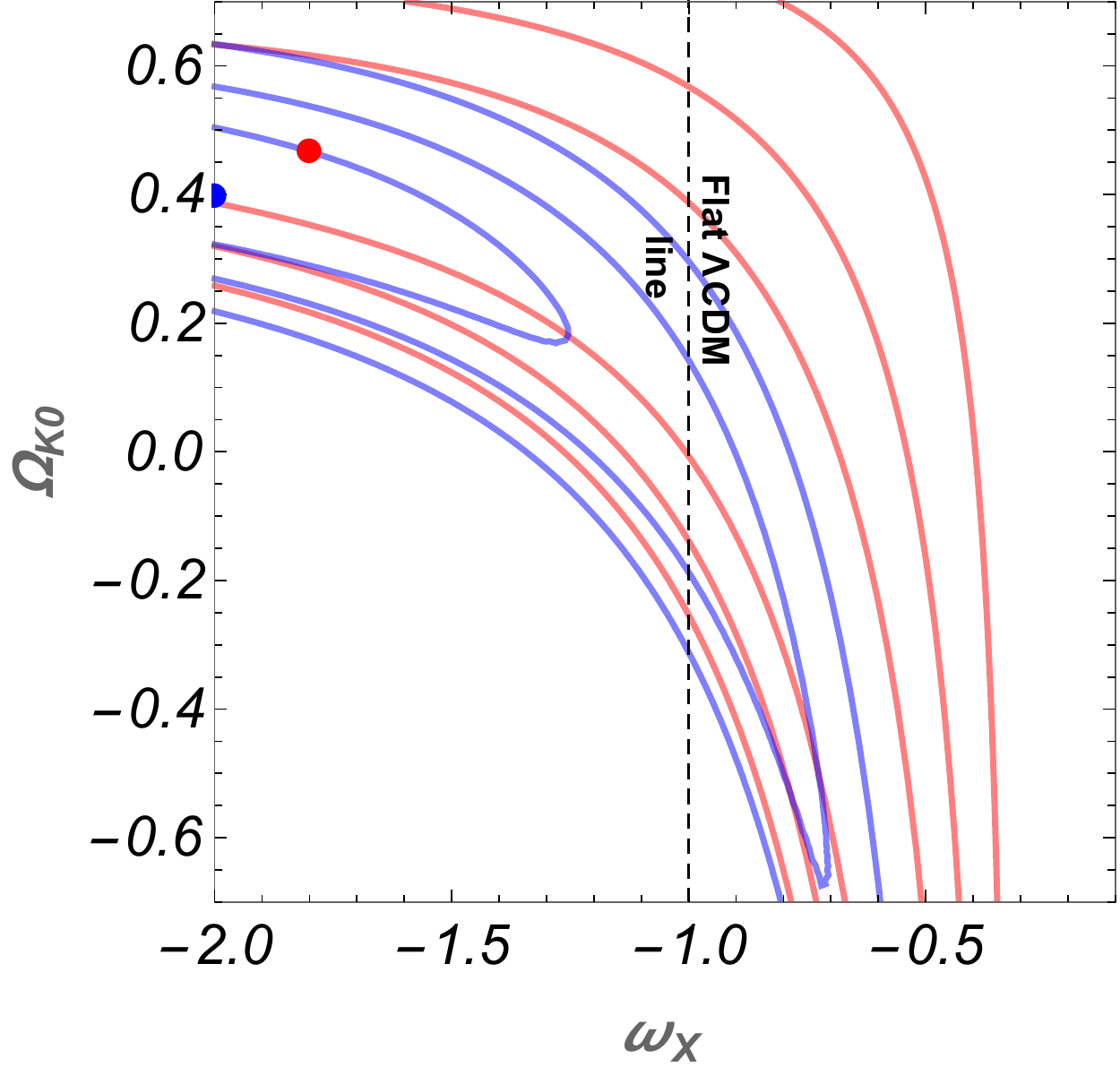}
\caption{
The three panels (from left to right) show $1\sigma$, $2\sigma$, 
and $3\sigma$ two-dimensional constraint contours for the three-parameter, 
non-flat XCDM parameterization, computed after marginalizing over each 
of the three parameters in turn. Red (blue) solid lines are for the 
lower (higher) $H_0$ prior. Left, center, and right panels correspond 
to marginalizing over $\Omega_{K0}$, $\omega_X$, and $\Omega_{m0}$ 
respectively. Red (blue) solid circles are the best-fit points for the 
lower (higher) $H_0$ prior. Red (blue) dot-dashed lines in the left 
panel are $1\sigma$, $2\sigma$, and $3\sigma$ constraint contours 
for the lower (higher) $H_0$ prior for spatially-flat XCDM (see the central panel 
of Fig.\ \ref{fig:Flatmodelsconstraints}). For 
quantitative parameter best-fit values and ranges see Tables 
\ref{table:Un-binned data zda}, \ref{table:Nonflatmodels}, and 
\ref{table:1Dfor3parametermodels}.
} \label{fig:NonflatXCDMconstraints}
\end{figure}

\begin{figure}[H]
\centering
  \includegraphics[width=54.8mm]{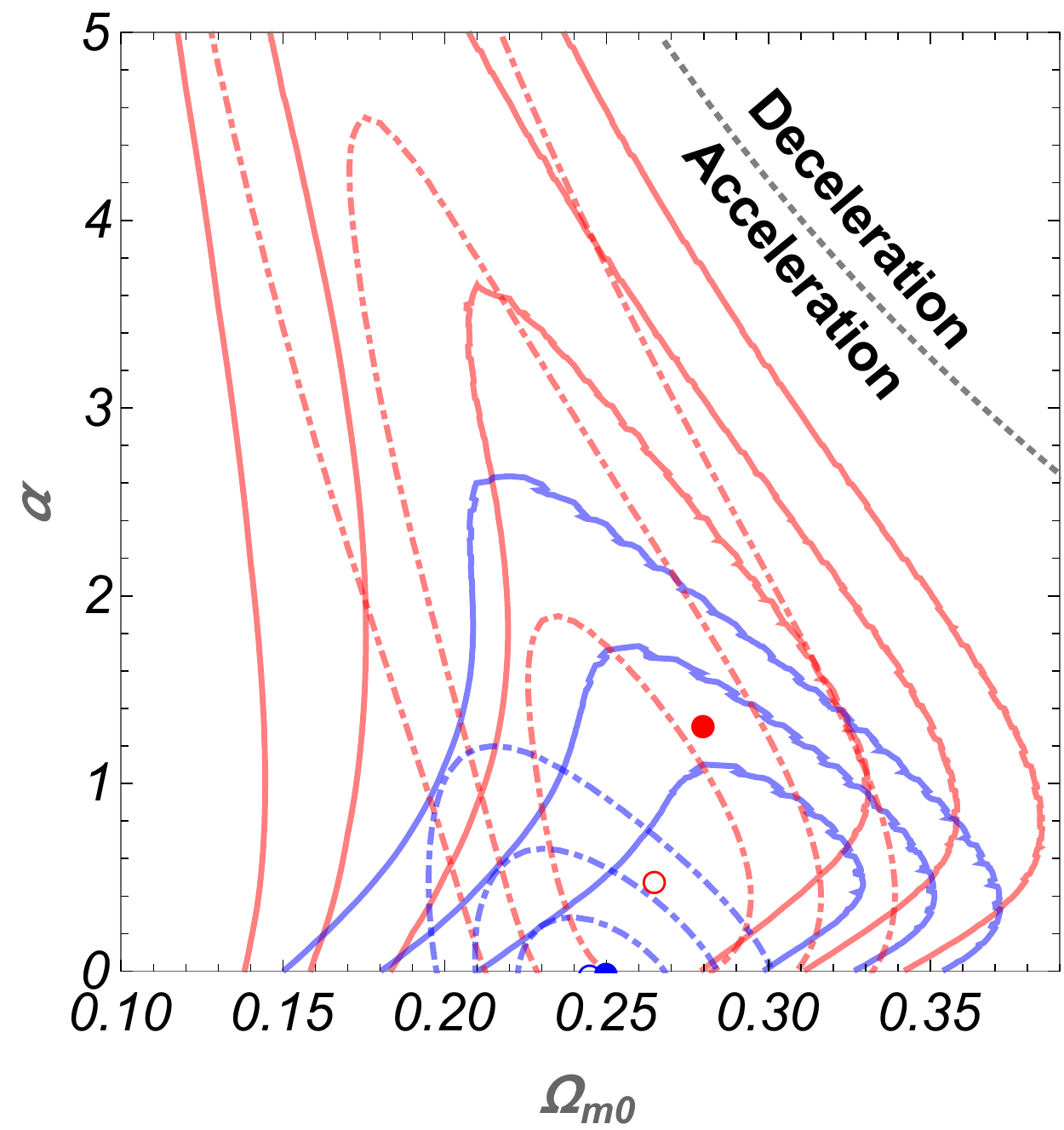}
  \includegraphics[width=62.0mm]{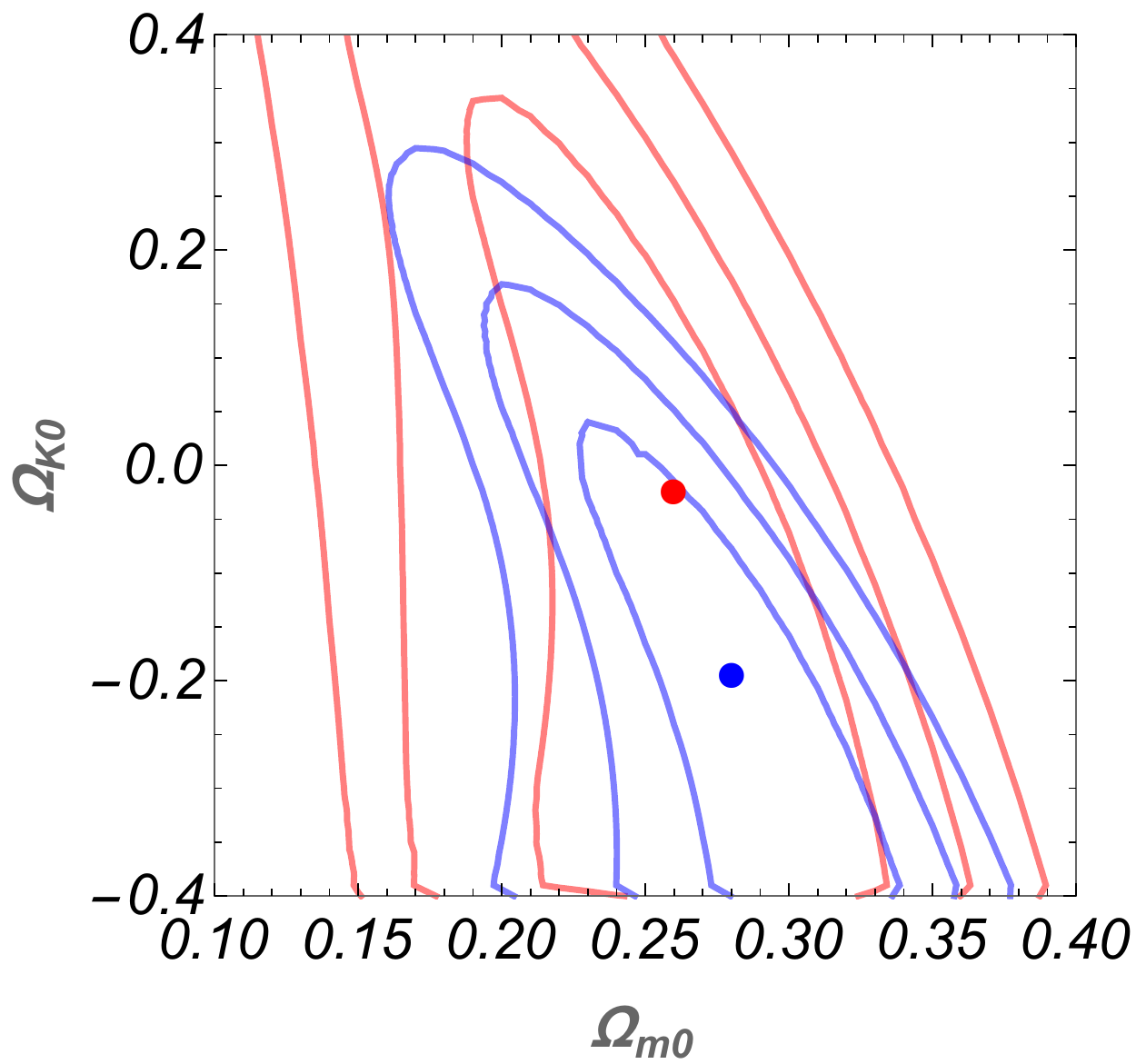}
  \includegraphics[width=61.4mm]{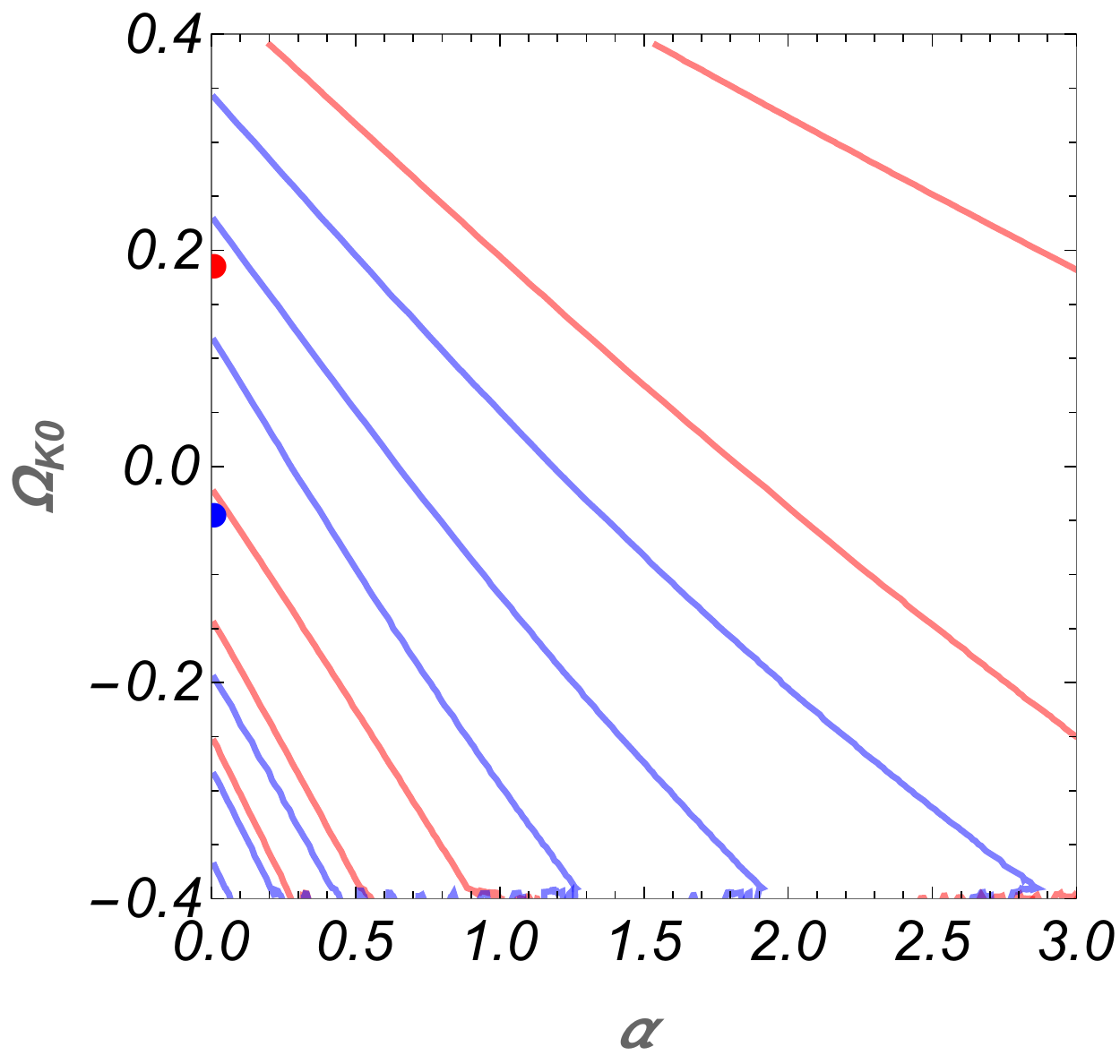}
\caption{
The three panels (from left to right) show $1\sigma$, $2\sigma$, and $3\sigma$ 
two-dimensional constraint contours for the three-parameter, 
non-flat $\phi$CDM model, computed after marginalizing over each 
of the three parameters in turn. Red (blue) solid lines are for the 
lower (higher) $H_{0}$ prior. Left, center, and right panels correspond to 
marginalizing over $\Omega_{K0}$, $\alpha$, and $\Omega_{m0}$ respectively. 
Red (blue) solid circles are the best-fit points for the lower (higher) $H_{0}$ 
prior. Red (blue) dot-dashed lines in the left panel are $1\sigma$, $2\sigma$,
and $3\sigma$ constraint contours for the lower (higher) $H_0$ prior 
for the spatially-flat $\phi$CDM model (see the right panel of Fig.\ 
\ref{fig:Flatmodelsconstraints}). For quantitative parameter best-fit 
values and ranges see Tables \ref{table:Un-binned data zda}, 
\ref{table:Nonflatmodels}, and \ref{table:1Dfor3parametermodels}.
} 
\label{fig:NonflatphiCDMconstraints}
\end{figure}

For the three-parameter models we next compute three two-dimensional 
likelihood functions by marginalizing the three-dimensional likelihood 
function over each of the three parameters (assuming flat priors) in turn.
These three two-dimensional likelihood functions are maximized as above and 
the corresponding best-fit parameter values and minimum $\chi^2$ are 
listed in Table\ \ref{table:Nonflatmodels}. The confidence contours 
for these two-dimensional likelihood functions are shown in Fig.\ 
\ref{fig:NonflatXCDMconstraints} for the non-flat XCDM parametrization 
and in Fig.\ \ref{fig:NonflatphiCDMconstraints} for the non-flat 
$\phi$CDM model.

To get two one-dimensional likelihood functions from each of the 
two-dimensional likelihood functions, we marginalize (with a flat
prior) over each parameter in turn. We then determine the best-fit 
parameter values by maximizing each one-dimensional likelihood 
function and compute  $1\sigma$, and $2\sigma$ intervals for
each parameter in each model and for both $H_0$ priors. The best-fit
parameter values and $1\sigma$ and $2\sigma$ intervals for the 
two-parameter models are given in Table \ref{table:1Dfor2parametermodels} 
and for the three-parameter models, these are given in Table 
\ref{table:1Dfor3parametermodels}.


The best-fit (two- and three-dimensional) model predictions are 
shown in Fig.\ \ref{fig:Error Plots}, for the five different 
cosmological models, $\Lambda$CDM in red, flat XCDM in blue, 
flat $\phi$CDM in green, non-flat XCDM in orange, 
and non-flat $\phi$CDM in brown, for the two $H_0$ priors, with $H_0 \pm \sigma_{H_0}=68\pm 2.8$ km s$^{-1}$ Mpc$^{-1}$ in solid lines and $H_0 \pm \sigma_{H_0}=73.24\pm 1.74$ km s$^{-1}$ Mpc$^{-1}$ in dot-dashed lines.


\begin{table*}[!htbp]

\small

\caption{Two-dimensional best-fit parameters for three-parameter, non-flat models}

\begin{center}

\begin{threeparttable}

\centering

\begin{tabular*}{\textwidth}{l@{\extracolsep{\fill}}ccccc}

\hline\hline 

\noalign{\vskip 1mm} 

\multirow{2}{*}{Model} & \multirow{2}{*}{$h$ Prior\tnote{a}} 

& \multirow{2}{*}{Marginalized Parameter}

& \multirow{2}{*}{BF\tnote{b}} 

& \multirow{2}{*}{$\chi_{\rm min}^2$} \\

{}&{}&{}&{}&{} \\

\hline \\*[-6pt]

\multirow{13}{*}{Non-flat XCDM} & \multirow{6}{*}{$0.68\pm 0.028$} & \multirow{2}{*}{$\Omega_{K0}$} & $\Omega_{m0}=0.38$ & \multirow{2}{*}{25.3} \\

{} & {} & {} & $\omega_{X}=-0.64$ & {} \\*[-4pt] \\*[-4pt]

{} & {} & \multirow{2}{*}{$\omega_{X}$} & $\Omega_{m0}=0.16$ & \multirow{2}{*}{22.5} \\

{} & {} & {} & $\Omega_{K0}=0.43$ & {} \\*[-4pt] \\*[-4pt]

{} & {} & \multirow{2}{*}{$\Omega_{m0}$} & $\omega_{X}=-1.80$ & \multirow{2}{*}{27.3} \\

{} & {} & {} & $\Omega_{K0}=0.47$ & {} \\*[-4pt] \\*[-4pt]

\cline{2-5} \\*[-6pt]

{} & \multirow{6}{*}{$0.7324\pm 0.0174$} & \multirow{2}{*}{$\Omega_{K0}$} & $\Omega_{m0}=0.13$ & \multirow{2}{*}{25.0} \\

{} & {} & {} & $\omega_{X}=-2$ & {} \\*[-4pt] \\*[-4pt]

{} & {} & \multirow{2}{*}{$\omega_{X}$} & $\Omega_{m0}=0.15$ & \multirow{2}{*}{22.1} \\

{} & {} & {} & $\Omega_{K0}=0.39$ & {} \\*[-4pt] \\*[-4pt]

{} & {} & \multirow{2}{*}{$\Omega_{m0}$} & $\omega_{X}=-2$ & \multirow{2}{*}{26.8} \\

{} & {} & {} & $\Omega_{K0}=0.41$ & {} \\*[-4pt] \\*[-4pt]

\hline \\*[-6pt]

\multirow{13}{*}{Non-flat $\phi$CDM} & \multirow{6}{*}{$0.68\pm 0.028$} & \multirow{2}{*}{$\Omega_{K0}$} & $\Omega_{m0}=0.28$ & \multirow{2}{*}{25.7} \\

{} & {} & {} & $\alpha=1.33$ & {} \\*[-4pt] \\*[-4pt]

{} & {} & \multirow{2}{*}{$\alpha$} & $\Omega_{m0}=0.26$ & \multirow{2}{*}{22.4} \\

{} & {} & {} & $\Omega_{K0}=-0.02$ & {} \\*[-4pt] \\*[-4pt]

{} & {} & \multirow{2}{*}{$\Omega_{m0}$} & $\alpha=0.01$ & \multirow{2}{*}{28.7} \\

{} & {} & {} & $\Omega_{K0}=0.19$ & {} \\*[-4pt] \\*[-4pt]

\cline{2-5} \\*[-6pt]

{} & \multirow{6}{*}{$0.7324\pm 0.0174$} & \multirow{2}{*}{$\Omega_{K0}$} & $\Omega_{m0}=0.25$ & \multirow{2}{*}{29.0} \\

{} & {} & {} & $\alpha=0.01$ & {} \\*[-4pt] \\*[-4pt]

{} & {} & \multirow{2}{*}{$\alpha$} & $\Omega_{m0}=0.28$ & \multirow{2}{*}{26.6} \\

{} & {} & {} & $\Omega_{K0}=-0.19$ & {} \\*[-4pt] \\*[-4pt]

{} & {} & \multirow{2}{*}{$\Omega_{m0}$} & $\alpha=0.01$ & \multirow{2}{*}{31.6} \\

{} & {} & {} & $\Omega_{K0}=-0.04$ & {} \\*[-4pt] \\*[-4pt]

\hline  

\hline

\end{tabular*}

\begin{tablenotes}

\item[a] Hubble constant in units of 100 km s$^{-1}$ Mpc$^{-1}$.

\item[b] Best-fit parameter values.

\end{tablenotes}

\end{threeparttable}

\end{center} 

\label{table:Nonflatmodels}

\end{table*}




While the main purpose of our paper was to improve on the characterization
of the deceleration-acceleration transition studied in \cite{FarooqRatra2013b} 
and \cite{Farooqetal2013b}, we see from Fig.\ \ref{fig:Flatmodelsconstraints}
and the left panels of Figs.\ \ref{fig:NonflatXCDMconstraints} 
and \ref{fig:NonflatphiCDMconstraints} that the $H(z)$ data by themselves
indicate that the cosmological expansion is currently accelerating.

From these figures, it is clear that the $H(z)$ data of Table
\ref{table:Hzdata} are very consistent with the standard spatially-flat 
$\Lambda$CDM cosmological model, although even for the two-parameter 
model constraint contours shown in Fig.\ \ref{fig:Flatmodelsconstraints}
there is a large range of dynamical dark energy models as well as 
spatially-curved models that are consistent with the data. In Figs.\ 
\ref{fig:NonflatXCDMconstraints} and \ref{fig:NonflatphiCDMconstraints} for 
the non-flat dynamical dark energy models, it is clear that allowing for 
non-zero space curvature considerably broadens the dynamical dark energy 
options and vice versa. It is interesting to note that in the non-flat
$\phi$CDM model \cite{Chenetal2016b} find that the cosmological data bound 
on the sum of neutrino masses is considerably weaker than if the model
were spatially flat.

While the error bars are large, it is curious that Table 
\ref{table:1Dfor3parametermodels} entries show that the non-flat XCDM
parametrization mildly favors open spatial hypersurfaces while the non-flat
$\phi$CDM model mildly prefers closed ones.

\begin{table*}[!htbp]
\small
\caption{One-dimensional best-fit parameters and intervals for two-parameter models}
\begin{center}
\begin{threeparttable}
\centering
\begin{tabular*}{\textwidth}{l@{\extracolsep{\fill}}cccccc}
\hline\hline 
\noalign{\vskip 1mm}

\multirow{2}{*}{Model} & \multirow{2}{*}{$h$ Prior\tnote{a}} & Marginalization & \multirow{2}{*}{BF\tnote{b}} & \multirow{2}{*}{1$\sigma$ intervals} & \multirow{2}{*}{2$\sigma$ intervals } \\

\noalign{\vskip 2mm} 

{} & {}  &  {Range} & {}& {} & {}\\

\noalign{\vskip 1mm} 

\hline

\multirow{8}{*}{$\Lambda$CDM} & \multirow{4}{*}{0.68 $\pm$ 0.028} & \multirow{2}{*}{$0 \leq \Omega_{\Lambda} \leq 1.4$} & \multirow{2}{*}{ 0.23} &  \multirow{2}{*}{$0.19 \leq \Omega_{m0} \leq 0.27$} & \multirow{2}{*}{$0.15 \leq \Omega_{m0} \leq 0.30$} \\

&&&&&\\


{}&{}& \multirow{2}{*}{$0 \leq \Omega_{m0} \leq 1$} &\multirow{2}{*}{ 0.58} &  \multirow{2}{*}{$0.46 \leq \Omega_{\Lambda} \leq 0.69$} & \multirow{2}{*}{$0.32 \leq \Omega_{\Lambda} \leq 0.80$} \\ 

&&&&&\\

\cline{2-6}

{} & \multirow{4}{*}{$0.7324 \pm 0.0174$} & \multirow{2}{*}{$0 \leq \Omega_{\Lambda} \leq 1.4$} &\multirow{2}{*}{0.26 } &  \multirow{2}{*}{$0.22 \leq \Omega_{m0} \leq 0.29$} & \multirow{2}{*}{$0.19 \leq \Omega_{m0} \leq 0.32$} \\

&&&&&\\


{}&{}& \multirow{2}{*}{$0 \leq \Omega_{m0} \leq 1$} &\multirow{2}{*}{0.79 } &  \multirow{2}{*}{$0.71 \leq \Omega_{\Lambda} \leq 0.86$} & \multirow{2}{*}{$0.63 \leq \Omega_{\Lambda} \leq 0.93$} \\

&&&&&\\

\hline

\multirow{8}{*}{Flat XCDM} & \multirow{4}{*}{0.68 $\pm$ 0.028} & \multirow{2}{*}{$-2 \leq \omega_{X} \leq 0$} &\multirow{2}{*}{ 0.27}&  \multirow{2}{*}{$0.25 \leq \Omega_{m0} \leq 0.29$} & \multirow{2}{*}{$0.22 \leq \Omega_{m0} \leq 0.31$} \\

&&&&&\\


{}&{}& \multirow{2}{*}{$0 \leq \Omega_{m0} \leq 1$} &\multirow{2}{*}{ $-0.85$} &  \multirow{2}{*}{$-0.98 \leq \omega_{X} \leq -0.73$} & \multirow{2}{*}{$-1.11 \leq \omega_{X} \leq -0.59$} \\

&&&&&\\

\cline{2-6}

{} & \multirow{4}{*}{$0.7324 \pm 0.0174$} & \multirow{2}{*}{$-2 \leq \omega_{X} \leq 0$} &\multirow{2}{*}{0.25 } &  \multirow{2}{*}{$0.23 \leq \Omega_{m0} \leq 0.26$} & \multirow{2}{*}{$0.22 \leq \Omega_{m0} \leq 0.28$} \\

&&&&&\\


{}&{}& \multirow{2}{*}{$0 \leq \Omega_{m0} \leq 1$} &\multirow{2}{*}{$-1.07$ } &  \multirow{2}{*}{$-1.17 \leq \omega_{X} \leq -0.98$} & \multirow{2}{*}{$-1.27 \leq \omega_{X} \leq -0.89$} \\

&&&&&\\

\hline

\multirow{8}{*}{Flat $\phi$CDM} & \multirow{4}{*}{0.68 $\pm$ 0.028} & \multirow{2}{*}{$0 \leq \alpha \leq 5$}  &\multirow{2}{*}{ 0.26} &\multirow{2}{*}{$0.23 \leq \Omega_{m0} \leq 0.28$} & \multirow{2}{*}{$0.20 \leq \Omega_{m0} \leq 0.30$} \\

&&&&&\\


{}&{}& \multirow{2}{*}{$0 \leq \Omega_{m0} \leq 1$} &\multirow{2}{*}{0.53 }&  \multirow{2}{*}{$0.09 \leq \alpha \leq 1.29$} & \multirow{2}{*}{$0 \leq \alpha \leq 2.9$} \\ 

&&&&&\\

\cline{2-6}

{} & \multirow{4}{*}{$0.7324 \pm 0.0174$} & \multirow{2}{*}{$0 \leq \alpha \leq 5$} &\multirow{2}{*}{ 0.24} &  \multirow{2}{*}{$0.23 \leq \Omega_{m0} \leq 0.26$} & \multirow{2}{*}{$0.21 \leq \Omega_{m0} \leq 0.28$} \\

&&&&&\\


{}&{}& \multirow{2}{*}{$0 \leq \Omega_{m0} \leq 1$} &\multirow{2}{*}{ 0} &  \multirow{2}{*}{$0 \leq \alpha \leq 0.15$} & \multirow{2}{*}{$0 \leq \alpha \leq 0.46$} \\

&&&&&\\

\hline  

\hline

\end{tabular*}

\begin{tablenotes}

\item[a] Hubble constant in units of 100 km s$^{-1}$ Mpc$^{-1}$.

\item[b] Best-fit parameter values.

\end{tablenotes}

\end{threeparttable}

\end{center} 

\label{table:1Dfor2parametermodels}

\end{table*}



\begin{table*}[!htbp]
\small
\caption{One-dimensional best-fit parameters and intervals for three-parameter, non-flat models}
\begin{center}
\begin{threeparttable}

\centering

\begin{tabular*}{\textwidth}{l@{\extracolsep{\fill}}cccccc}

\hline\hline 

\noalign{\vskip 1mm} 

\multirow{2}{*}{Model} & \multirow{2}{*}{$h$ Prior\tnote{a}} & Marginalization  & \multirow{2}{*}{BF} &\multirow{2}{*}{1$\sigma$ intervals} & \multirow{2}{*}{2$\sigma$ intervals} \\

\noalign{\vskip 2mm} 

{} & {}  &  {Range\tnote{b}} & {}& {} & {}\\

\noalign{\vskip 1mm} 

\hline \\*[-6pt]

\multirow{14}{*}{Non-flat XCDM} & \multirow{6}{*}{0.68 $\pm$ 0.028} & \multirow{1}{*}{$0 \leq \Omega_{m0} \leq 1$} &  \multirow{2}{*}{ 0.45}  &  \multirow{2}{*}{$0.32 \leq \Omega_{K0} \leq 0.55$} & \multirow{2}{*}{$-0.06 \leq \Omega_{K0} \leq 0.66$} \\

{}&{}& $-2 \leq \omega_X \leq 0$ & & &\\*[-4pt]\\*[-2pt]


{}&{}& \multirow{1}{*}{$0 \leq \Omega_{m0} \leq 1$} &  \multirow{2}{*}{$-0.71$}&  \multirow{2}{*}{$-1.28 \leq \omega_X \leq -0.59$} & \multirow{2}{*}{$-2 \leq \omega_X \leq -0.49$} \\

{}&{}& $-0.7 \leq \Omega_{K0} \leq 0.7$ & & &\\*[-4pt]\\*[-2pt]

{}&{}& \multirow{1}{*}{$-2 \leq \omega_{X} \leq 0$} &  \multirow{2}{*}{0.26} &  \multirow{2}{*}{$0.21 \leq \Omega_{m0} \leq 0.33$} & \multirow{2}{*}{$0.16 \leq \Omega_{m0} \leq 0.39$} \\

{}&{}& $-0.7 \leq \Omega_{K0} \leq 0.7$ & & &\\*[-4pt]\\*[-2pt]

\cline{2-6} \\*[-6pt]

{} & \multirow{6}{*}{$0.7324 \pm 0.0174$} & \multirow{1}{*}{$0 \leq \Omega_{m0} \leq 1$} &  \multirow{2}{*}{0.36} &  \multirow{2}{*}{$0.28 \leq \Omega_{K0} \leq 0.43$} & \multirow{2}{*}{$0.12 \leq \Omega_{K0} \leq 0.50$} \\

{}&{}& $-2 \leq \omega_X \leq 0$ & & &\\*[-4pt]\\*[-2pt]


{}&{}& \multirow{1}{*}{$0 \leq \Omega_{m0} \leq 1$} &  \multirow{2}{*}{$-0.98$ } &  \multirow{2}{*}{$-1.54 \leq \omega_X \leq -0.91$} & \multirow{2}{*}{$-2 \leq \omega_X \leq -0.70$} \\

{}&{}& $-0.7 \leq \Omega_{K0} \leq 0.7$ & & &\\*[-4pt]\\*[-2pt]

{}&{}& \multirow{1}{*}{$-2 \leq \omega_{X} \leq 0$} &  \multirow{2}{*}{0.22} &  \multirow{2}{*}{$0.18 \leq \Omega_{m0} \leq 0.26$} & \multirow{2}{*}{$0.10 \leq \Omega_{m0} \leq 0.35$} \\

{}&{}& $-0.7 \leq \Omega_{K0} \leq 0.7$ & & &\\*[-4pt]\\*[-2pt]

\hline \\*[-6pt]

\multirow{14}{*}{Non-flat $\phi$CDM } & \multirow{6}{*}{0.68 $\pm$ 0.028} & \multirow{1}{*}{$0 \leq \Omega_{m0} \leq 1$} &  \multirow{2}{*}{$-0.28$} &  \multirow{2}{*}{$-0.4 \leq \Omega_{K0} \leq 0.10$} & \multirow{2}{*}{$-0.4 \leq \Omega_{K0} \leq 0.38$} \\

{}&{}& $0 \leq \alpha \leq 5$ & & &\\*[-4pt]\\*[-2pt]


{}&{}& \multirow{1}{*}{$0 \leq \Omega_{m0} \leq 1$} &  \multirow{2}{*}{0.087} &  \multirow{2}{*}{$0 \leq \alpha \leq 2.03$} & \multirow{2}{*}{$0 \leq \alpha \leq 4.07$} \\

{}&{}& $-0.4 \leq \Omega_{K0} \leq 0.4$ & & &\\*[-4pt]\\*[-2pt]

{}&{}& \multirow{1}{*}{$0 \leq \alpha \leq 5$} &  \multirow{2}{*}{ 0.26} &  \multirow{2}{*}{$0.21 \leq \Omega_{m0} \leq 0.30$} & \multirow{2}{*}{$0.17 \leq \Omega_{m0} \leq 0.33$} \\

{}&{}& $-0.4 \leq \Omega_{K0} \leq 0.4$ & & &\\*[-4pt]\\*[-2pt]

\cline{2-6} \\*[-6pt]

{} & \multirow{6}{*}{$0.7324 \pm 0.0174$} & \multirow{1}{*}{$0 \leq \Omega_{m0} \leq 1$} &  \multirow{2}{*}{$-0.35$} &  \multirow{2}{*}{$-0.4 \leq \Omega_{K0} \leq -0.08$} & \multirow{2}{*}{$-0.4 \leq \Omega_{K0} \leq 0.08$} \\

{}&{}& $0 \leq \alpha \leq 5$ & & &\\*[-4pt]\\*[-2pt]


{}&{}& \multirow{1}{*}{$0 \leq \Omega_{m0} \leq 1$} &  \multirow{2}{*}{ 0} &  \multirow{2}{*}{$0 \leq \alpha \leq 0.71$} & \multirow{2}{*}{$0 \leq \alpha \leq 1.25$} \\

{}&{}& $-0.4 \leq \Omega_{K0} \leq 0.4$ & & &\\*[-4pt]\\*[-2pt]

{}&{}& \multirow{1}{*}{$0 \leq \alpha \leq 5$} &  \multirow{2}{*}{0.28} &  \multirow{2}{*}{$0.25 \leq \Omega_{m0} \leq 0.31$} & \multirow{2}{*}{$0.22 \leq \Omega_{m0} \leq 0.34$} \\

{}&{}& $-0.4 \leq \Omega_{K0} \leq 0.4$ & & &\\*[-4pt]\\*[-2pt]

\hline  
\hline

\end{tabular*}
\begin{tablenotes}
\item[a] Hubble constant in units of 100 km s$^{-1}$ Mpc$^{-1}$.
\item[b] The three-parameter-dependent likelihood function is integrated
over the two parameters in the ranges given in this column and the 
corresponding best-fit and 1$\sigma$ and 2$\sigma$ intervals of the 
third parameter is computed and listed in the fourth, fifth, and sixth columns 
respectively of the same row.
\end{tablenotes}
\end{threeparttable}
\end{center} 
\label{table:1Dfor3parametermodels}
\end{table*}


\section{Conclusion}

From the new list of $H(z)$ data we have compiled, we find evidence for the 
cosmological deceleration-acceleration transition to have taken place at 
a redshift $z_{\rm da} = 0.72 \pm 0.05\ (0.84 \pm 0.03)$, depending on the 
value of  $H_0 = 68 \pm 2.8\ (73.24 \pm 1.74)$ km s$^{-1}$ Mpc$^{-1}$, but
otherwise only mildly dependent on other cosmological parameters. In addition, the binned $H(z)$ data in redshift space show qualitative visual 
evidence for the deceleration-acceleration transition, independent of 
how they are binned provided the bins are narrow enough, in agreement with that originally found by \cite{Farooqetal2013b}.
These $H(z)$ data are consistent with the standard spatially-flat 
$\Lambda$CDM cosmological model but do not rule out non-zero space
curvature or dynamical dark energy, especially in models that allow for
both. Other data, such as currently available SNIa, BAO, growth factor, 
or CMB anisotropy data can tighten the constraints on these parameters
\cite[see, e.g.][]{Farooqetal2015}, and it is of interest to study 
how the other data constrains parameters when used in conjunction with 
the $H(z)$ data we have compiled here.

\section*{Acknowledgements}

O.F.\ and M.F. greatfully acknowledge the partial funding from the Department of Physical Sciences, Embry-Riddle Aeronautical University. S.C.\ and B.R.\ were supported in part by DOE grant DE-SC0011840.


\newpage

\end{document}